\documentclass[twocolumn]{aastex631}
\usepackage{tabularx,ragged2e} 
\usepackage{multirow}
\usepackage{booktabs}
\usepackage{amsmath}
\usepackage{comment}
\submitjournal{ApJ}

\begin{document}

\title{Improving Solar EUV Irradiance Modeling with Differential Emission Measure Informed Spectra}

\newcommand{\lasp}{
	Laboratory for Atmospheric and Space Physics,
	University of Colorado at Boulder,
	Boulder, CO, USA
	}

 \newcommand{\cires}{
	Cooperative Institute for Research in Environmental Sciences,
	University of Colorado,
	Boulder, CO USA
	}
	
\newcommand{\ncei}{
	NOAA National Centers for Environmental Information,
	Boulder, CO USA
	}

 \newcommand{\swri}{
	Southwest Research Institute,
	Boulder, CO, USA
	}

  \newcommand{\lm}{
	Lockheed Martin Advanced Technology Center,
    Palo Alto, CA, USA
	}

  \newcommand{\cisro}{
	Commonwealth Scientific and Industrial Research Organisation, 
    Canberra, Australian Capital Territory, AU
    }

\author[0000-0002-7586-4220]{Courtney L. Peck}
\affiliation{\lasp}
\affiliation{\cires}
\affiliation{\ncei}

\author[0000-0003-0016-5377]{Shah M. Bahauddin}
\affiliation{\lasp}

\author[0000-0002-0494-2025]{Daniel B. Seaton}
\affiliation{\swri}

\author[0000-0002-0344-0314]{Janet L. Machol}
\affiliation{\cires}
\affiliation{\ncei}


\author[0000-0002-5305-9466]{Ed Thiemann}
\affiliation{\lasp}

\author[0000-0003-2110-9753]{Mark Cheung}
\affiliation{\lm}
\affiliation{\cisro}

\begin{abstract}
Solar extreme ultraviolet (EUV) irradiance predominantly originates from the chromosphere and corona, and variations in EUV irradiance are a significant driver of space weather at Earth, causing increased satellite drag, radio communication disruptions, and reduced GPS accuracy. Since solar EUV irradiance measurements are intermittent and inhomogeneous, models are used to fill spectral and temporal gaps, typically relying on proxy-based methods that utilize linear correlations between emission originating from similar heights or temperatures in the solar atmosphere. Proxy methods can struggle to capture dynamic irradiance variations of the solar corona during flares in part due to the relative infrequency of flares on which to derive correlations and the steep temporal temperature gradients experienced during flares. This paper presents a hybrid method for modeling solar EUV irradiance utilizing forward-modeled physics-informed solar differential emission measures (DEMs) to capture the optically-thin coronal emission coupled with traditional proxy-based correlation methods to capture the chromospheric and continuum emission. The model is trained and tested using inputs from the Solar Dynamics Observatory (SDO) Atmospheric Imaging Assembly (AIA) and EUV Variability Experiment (EVE) and compared against a proxy-only method. The results demonstrate that including information from DEMs into a EUV irradiance model improves accuracy for flaring conditions, particularly for high-temperature coronal lines, where the model error is reduced by up to a factor of 3 for M- and X-class flares compared to a proxy-only model trained on the same data.

\end{abstract}


\section{Introduction} \label{sec:intro}
Solar X-ray and extreme ultraviolet (EUV) irradiance varies by a factor of 2 to more than 10 on timescales from minutes to the 11-year solar cycle \citep[see e.g.,][for reviews]{lilensten_review_2008, kretzschmar_solar_2009}. This variability constitutes a fundamental driver of energy input into Earth’s upper atmosphere and plays a central role in regulating the thermodynamic state of the thermosphere and ionosphere. While geomagnetic forcing associated with high-speed solar wind streams and coronal mass ejections can strongly perturb the thermospheric density \citep[e.g.,][]{chen_discrepancy_2012, mcgranaghan_impact_2014}, solar EUV irradiance remains the primary driver of thermospheric energy input at low to mid latitudes, even during geomagnetically active periods \citep{thiemann_goes-r_2019}. In this context, the atmospheric response is governed not only by the total EUV flux but also by its spectral distribution, since different wavelengths are absorbed at different altitudes. Capturing this vertical structuring of energy deposition, and the resulting temperature and density fluctuations, therefore requires accurate, high-resolution measurements or physically consistent models of solar spectral EUV irradiance coupled with thermosphere–ionosphere models. 

The density variations have immediate practical consequences for the near-Earth space environment. In low Earth orbit, atmospheric drag is the largest source of uncertainty in orbit determination and prediction, owing largely to the complex and highly variable thermospheric response to solar EUV forcing \citep{fedrizzi_global_2012}. Errors in density prediction propagate directly into atmospheric drag calculations, and as a result they introduce errors into orbit prediction and increase uncertainty in conjunction assessment and collision avoidance. \citep[e.g.,][]{emmert_thermospheric_2015,emmert_propagation_2017,vourlidas_euv_2018}.

High-resolution ($\gtrapprox$ 0.1 nm) solar EUV irradiance measurements provide powerful spectral diagnostics of plasma conditions in the chromosphere and corona \citep[see][for a review]{del_zanna_solar_2018}. These diagnostics enable studies of the temporal evolution of electron temperature, density, emission measure, and elemental abundances across a wide range of solar conditions \citep[e.g.,][]{2013SoPh..283....5A, kretzschmar_variability_2004, zanna_spectral_2013, aschwanden_global_2014, schonfeld_slowly_2017}. During solar flares, in particular, such measurements are routinely used to probe the temperature and density evolution of coronal plasma \citep[e.g.,][]{zanna_spectral_2013}, helping to answer questions about the mechanism of energy release and the impact on coronal heating. Moreover, because the solar EUV spectrum contains a high density of emission lines, high spectral resolution mitigates the effects of line blending and thus increases the available diagnostic information.

\begin{figure*}[ht!]
    \centering
    \includegraphics[width=0.86\textwidth, trim={10mm, 5mm, 5mm, 10mm}]{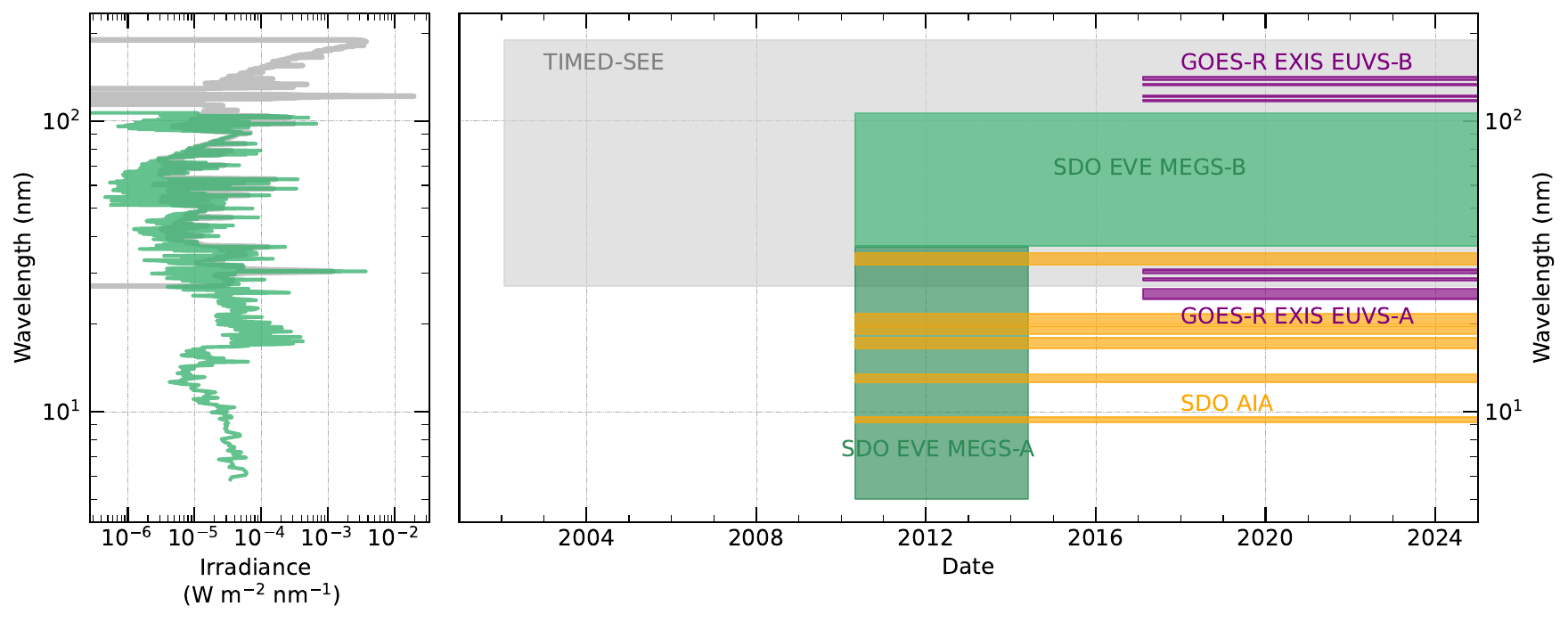}
    \vspace{-3mm}
    \caption{\textit{Left:} Spectra from the TIMED-SEE and EVE MEGS-A and -B instruments. \textit{Right:} Temporal coverage of TIMED-SEE and EVE MEGS-A and -B spectral observations, SDO AIA images, and GOES-R EXIS EUVS-A and -B narrow-band observations. EVE MEGS-A suffered a capacitor failure in 2014, preventing further measurements. }
    \label{fig:eve_timedsee}
\end{figure*}

Despite a long history of solar EUV measurements (from $\sim$5 to 127 nm), the temporal and spectral coverage of these irradiances has been intermittent and inhomogeneous. Figure~\ref{fig:eve_timedsee} summarizes the spectral and temporal coverage of two instruments that have provided robust, spectrally resolved solar EUV irradiance measurements: the Thermosphere Ionosphere Mesosphere Energetics Dynamics-Solar EUV Experiment \citep[TIMED-SEE;][]{woods_solar_2005} and the Solar Dynamics Observatory's Extreme Ultraviolet Variability Experiment \citep[EVE;][]{woods_extreme_2012}. TIMED-SEE measures solar EUV irradiance from 27 to 190~nm at 0.4~nm resolution; below 27~nm, only broadband diode measurements are available and the spectral distribution must be modeled. EVE measures solar spectral irradiance from 5 to 107~nm with 0.02~nm sampling using the two-channel Multiple EUV Grating Spectrographs (MEGS) instrument, with MEGS-A and MEGS-B measuring spectral irradiance from 5--37~nm and 37--107~nm, respectively. In May 2014, MEGS-A suffered a capacitor short that prevented further observations. In combination, the limited spectral resolution of TIMED-SEE at shorter EUV wavelengths and the loss of EVE/MEGS-A have resulted in the absence of continuous, spectrally resolved EUV irradiance measurements between 5 and 27~nm, thereby creating a substantial gap in high-resolution coverage at the shortest EUV wavelengths (outside of annual EVE rocket underflight calibration campaigns \cite{hock_extreme_nodate}). More recently, since 2017, observations from NOAA's Geostationary Operational Environmental Satellite-R (GOES-R) Extreme Ultraviolet and X-ray Irradiance Sensors (EXIS) EUV Extreme Ultraviolet Sensor \citep[EUVS:][]{eparvier_extreme_2009} have provided EUV irradiance measurements for seven solar lines between 25.6 and 140.5~nm, as well as for the Mg II h and k lines at 280~nm, partially restoring operational monitoring but without full spectral coverage.

Models of solar EUV irradiance are commonly used to fill spectral and temporal gaps in measurements and to provide cross-calibration between instruments. Most existing models are empirical and rely on proxy measurements in which a limited set of well-observed EUV lines or broadband indices are used, using empirically calibrated relationships derived from periods of overlapping observations, to reconstruct EUV spectral irradiance in wavelength ranges that are not continuously measured (e.g., FISM1 and FISM2: \citet{chamberlin_euv_2007, chamberlin_flare_2008, chamberlin_flare_2020}, SOLAR2000: \citet{tobiska_solar2000_2000}, EUVAC: \citet{richards_euvac_1994}, SER1: \citet{hinteregger_observational_1981}, NRLEUV: \citet{warren_nrleuv_2006}, GOES-R EUVS model: \citet{thiemann_goes-r_2019}). Because these proxy-based approaches infer spectral variability indirectly, their accuracy depends critically on the availability of proxies that sample a broad range of plasma temperatures in the solar atmosphere. This requirement is particularly stringent in the solar corona, where temperatures span from $\sim$0.5 to $>20$~MK and can evolve rapidly during highly impulsive solar flares. Under extreme conditions, proxy models can fail to capture the full coronal emission, likely reflecting the lack of suitable proxies that track variability at the hottest temperatures and shortest timescales.

While proxy-based irradiance models have proven highly effective for reconstructing solar EUV variability across a wide range of conditions, their formulation is inherently tied to the set of observables used as inputs and the empirical relationships calibrated from them. By construction, such models infer spectral variability indirectly from a linear combination of a small number of broadband or line-integrated measurements, and therefore do not explicitly encode how emission is distributed across temperature in the solar atmosphere. This approach is generally not critical when the plasma is close to thermal equilibrium and remains well sampled by available proxies. However, during flares and other transient events, the atmosphere can develop broad, rapidly evolving temperature distributions. In such cases, incorporating additional physically motivated information about the plasma thermodynamics and the underlying atomic transitions has the potential to further improve spectral reconstructions.

Differential Emission Measure (DEM) analysis \citep[e.g.,][]{golub_solar_2009, hannah_differential_2012, cheung_thermal_2015, plowman_fast_2020} provides a physically-motivated description of the coronal plasma by quantifying how much emitting material is present at each temperature along the line of sight. In this framework, the observed emission is interpreted as the sum of contributions from plasma spanning a range of temperatures, rather than being characterized by a small number of discrete thermal components. When coupled with atomic emission models, DEMs can be used in forward modeling to reconstruct the emergent EUV spectrum. Recent studies have demonstrated that solar EUV images combined with DEM analysis and forward modeling can successfully reproduce EUV irradiance to within 20\% using five disk-integrated spectral lines \citep{kretzschmar_retrieving_2006} and 10\% percent using six spatially-resolved EUV image channels \citep{szenicer_deep_2019}.

In this work, we build on these advances by presenting a new hybrid method for modeling solar EUV irradiance that combines forward-modeled spectra from DEMs to capture the optically-thin coronal emission with proxy method to capture the optically-thick chromospheric and continuum emission. Unlike standard proxy approaches, which rely on linear combinations of measurements sampling a limited number of discrete temperature regimes, our DEM-based method encodes the thermal distribution of the corona coupled with the atomic database of line transitions. Thus it provides a more physically meaningful representation of the ionization states and emission processes, particularly during impulsive heating events or under non-equilibrium plasma conditions such as those during solar flares, while retaining the strengths of proxy methods where DEM assumptions break down.

This paper is organized as follows. In Section~\ref{sec:methods}, we describe the hybrid modeling framework, including the formulation of the proxy-only and hybrid models and the framework of calculating forward-modeled DEM spectra. Section~\ref{sec:data} introduces the observational datasets used for training and validation, including AIA, EVE, and EUVS, and details the construction of the proxy inputs and DEM-derived spectra. In Section~\ref{sec:model}, we outline the model training strategy, data selection, and validation methodology, with additional details provided in the Appendix. Section~\ref{sec:result} presents the results, including comparisons between the hybrid and proxy-only models across different solar conditions and spectral lines. Finally, we summarize the main findings, discuss the limitations of the approach, and outline directions for future work.

\section{Methodology} \label{sec:methods}

The hybrid model builds upon traditional proxy modeling techniques to model EUV irradiance by incorporating the forward-modeled spectra from DEMs in order to better capture coronal emission. Traditional proxy models for solar EUV irradiance (e.g., FISM2, GOES-R) rely on measurements at a variety of wavelengths sampling different temperatures in the solar atmosphere (i.e., the proxy measurements) to train coefficients that relate these measurements to the irradiance at unmeasured wavelengths. While modern proxy models typically decompose the proxy measurements into different timescales (e.g., high-cadence and time averaged measurements), we use a simple proxy model in this work to limit the influence on timescale selections on the model performance. The simple proxy model equation using linear regression is 
\begin{equation}
\label{equation:proxy_model}
    I(\lambda) = \sum_n c^{proxy}_n(\lambda) \times I^{proxy}_n + d(\lambda),
\end{equation}
where $I(\lambda)$ is the modeled irradiance at a given wavelengths, $I^{proxy}_n$ is the ${n}$th measured proxy measurement, $c_n$ is the coefficient that weights the nth proxy irradiance to $I(\lambda)$, $n$ is the number of proxy measurements used in the proxy model, and $d$ is a constant offset. Linear regression is commonly used to determine the correlation coefficients $c^{proxy}_n$ and $d$ which finds the best fit of the coefficients for the proxy measurements, $I^{proxy}_n$, to reproduce the measured spectra. Note that many proxy models subtract a baseline spectrum from the left-hand side of Equation~\ref{equation:proxy_model} to remove the constant offset term, $d(\lambda)$.

\subsection{Hybrid Method Incorporating Forward-Modeled DEM Emission}
This hybrid method incorporates physics-based knowledge of the coronal emission using DEMs into the proxy modeling. For optically-thin emission from a plasma in thermodynamic equilibrium, the temperature distribution of the plasma along the line-of-sight is described by the differential emission measure as 
\begin{equation}
\label{equation:DEM_definition}
    DEM(T)dT = \int_0^\infty n_e(T)^2 dz,
\end{equation}
where $n_e(T)$ is the electron number density of plasma at temperature $T$ along the line of sight, $dz$. The line-of-sight irradiance measurement, $y$, of an optically-thin plasma described by $DEM(T)$ measured by an instrument with temperature response function $R_\lambda(T)$ is given by
\begin{equation}
\label{equation:line_of_sight_measurement}
    y_i = \int_{T}^{} R_\lambda(T) DEM(T) \,dT,
\end{equation}
where $R_\lambda(T)$ depends on the plasma properties (elemental abundances and temperature) and the wavelength-dependent instrument responsivity. Codes such as \citet{plowman_fast_2020, cheung_thermal_2015, hannah_differential_2012} are commonly used to perform an inversion on Equation~{\ref{equation:line_of_sight_measurement}} to solve for $DEM(T)$. 

Since the DEM provides information on the temperature and emission measure of the coronal plasma, this information can be input into an atomic spectral model such as CHIANTI \citep{dere_chianti_1997, del_zanna_atomic_2020} to calculate the emergent intensity from the plasma. CHIANTI contains software packages that allow the reconstruction of a spectrum from any set of parameters including DEM-derived emission measures and temperatures. The reliability of spectra generated from CHIANTI based on DEM-derived quantities has been demonstrated in the EUV \citep[e.g.,][]{kretzschmar_retrieving_2006,szenicer_deep_2019}, as well as X-ray wavelengths \citep{su_determination_2018}. The resulting spectra, $I^{DEM}(\lambda)$, found in Equation~\ref{equation:DEM_irradiance} represents the plasma emission captured by the DEM, where $\epsilon(\lambda, T)$ is the emissivity table derived from the CHIANTI database (e.g., top panel of Figure~\ref{fig:responsivity})
\begin{equation}
\label{equation:DEM_irradiance}
    I^{DEM}(\lambda) = DEM(T)dT \times \epsilon(\lambda, T).
\end{equation}
From Equation~\ref{equation:line_of_sight_measurement}, solving for $DEM(T)$ relies on $y$ containing emergent intensity from all depths along the line-of-sight for the emitting plasma; a condition only satisfied for optically-thin plasmas of the solar corona. Therefore, the hybrid model includes a proxy-model component using Equation~\ref{equation:proxy_model} to model the optically-thick continuum and chromospheric emission. Adding forward-modeled DEM irradiance, $I^{DEM}(\lambda)$, from Equation~\ref{equation:DEM_irradiance}, gives the hybrid model equation as
\begin{equation}
\label{equation:hybrid_model}
    \begin{aligned}
    I(\lambda) =& \sum_m c^{DEM}_m(\lambda) \times I^{DEM}(\lambda_m)\\ &+ \sum_n c^{proxy}_n(\lambda) \times I^{proxy}_n + d(\lambda),
    \end{aligned}
\end{equation}
where $m$ represents the number of wavelength bins in $I^{DEM}(\lambda)$. Note that the second term on the right-hand-side of Equation~\ref{equation:hybrid_model} is identical to the proxy-only model from Equation~\ref{equation:proxy_model}. In this work, all coefficients ($c^{DEM}_m(\lambda)$, $c^{proxy}_n(\lambda)$, and $d(\lambda)$) are be determined using linear regression.

\section{Model Input Data} \label{sec:data}

The hybrid model is trained and tested using Level~1 data from the Atmospheric Imaging Assembly (AIA; \citealt{lemen_atmospheric_2012}) and Level~2B data from the EUV Variability Experiment (EVE; \citealt{woods_extreme_2012,eve_l2b_data}) Multiple EUV Grating Spectrograph (MEGS; \citealt{crotser_sdo-eve_2004}), both onboard the Solar Dynamics Observatory (SDO; \citealt{pesnell_solar_2012}). Observations from the Thermosphere Ionosphere Mesosphere Energetics Dynamics–Solar EUV Experiment (TIMED-SEE; \citealt{woods_solar_2005}) are also used to simulate GOES-R EXIS EUVS measurements. The datasets span the period from May 2010 to July 2013, prior to the MEGS-A failure in 2014 and excluding an extended interval in 2013 when MEGS-B data were flagged for low quality.

The hybrid model has two input components, as defined in Equation~\ref{equation:hybrid_model}: the proxy term, $I^{\mathrm{proxy}}$, and the DEM term, $I^{\mathrm{DEM}}$. The $I^{\mathrm{proxy}}$ input, described in Section~\ref{sec:EVE}, represents continuum emission from chromosphere and transition-region, analogous to those observed by instruments such as EUVS. The $I^{\mathrm{DEM}}$ input is derived from disk-integrated AIA observations of six coronal channels and represents the optically-thin coronal contribution, as described in Section~\ref{sec:AIA}.

\subsection{$I^{proxy}_n$ from EVE and TIMED-SEE Measurements} \label{sec:EVE}

\begin{figure}[ht!]
    \centering
    \includegraphics[width=0.47\textwidth, trim={8mm, 10mm, 5mm, 2mm}]{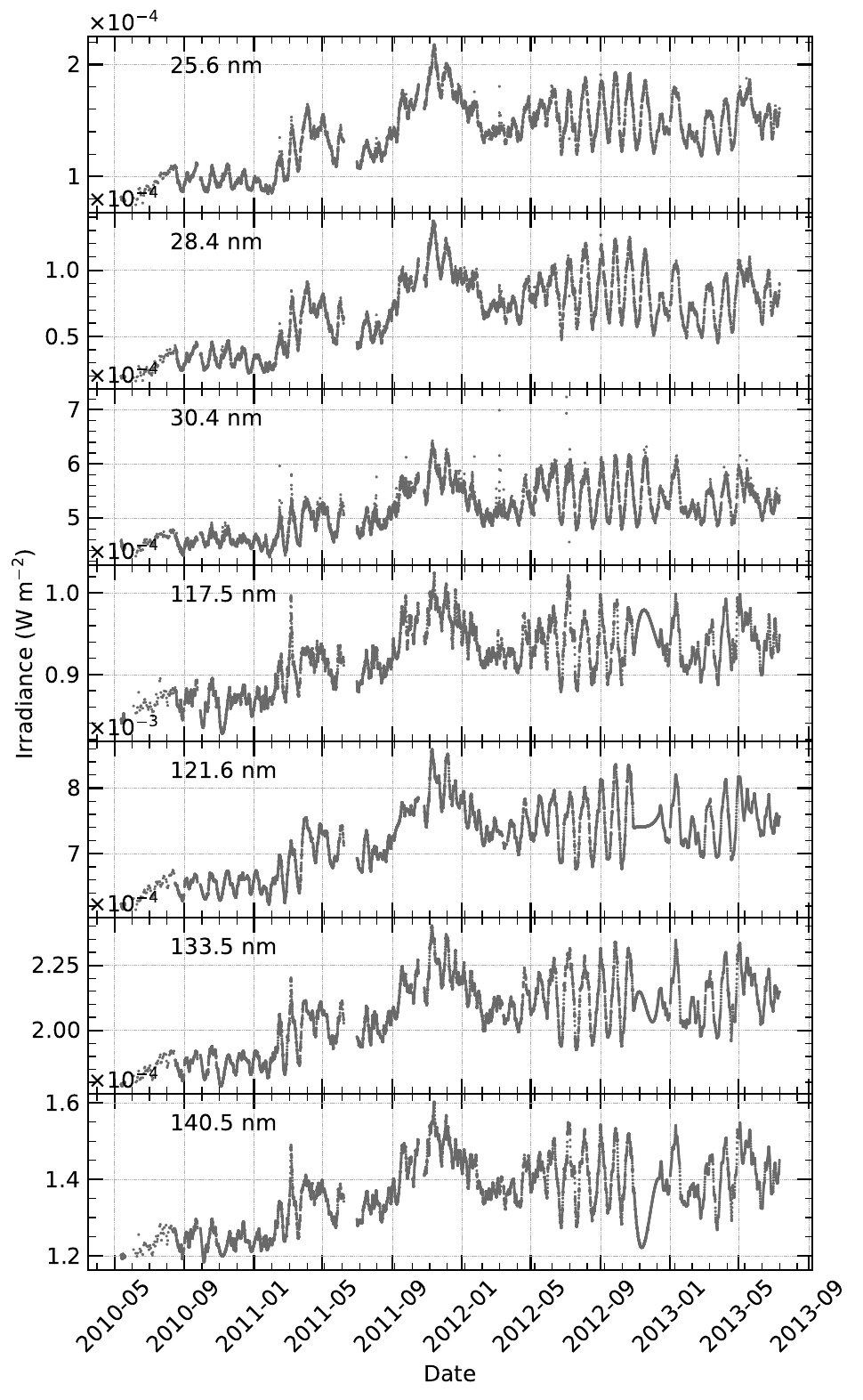}
    \vspace{-3mm}
    \caption{Simulated measurements for the 7 EUV spectral lines measured by the GOES-R EXIS EUVS instrument. The simulated measurements are created by convolving the EVE and TIMED-SEE spectra with the EUVS bandpasses.}
    \label{fig:Synthetic_EXIS}
\end{figure}

Simulated EUVS data at 60-second cadence are used as the proxy measurement inputs, $I^{\mathrm{proxy}}_n$. EUVS aboard GOES-R measures solar emission in seven discrete spectral bands (25.6, 28.4, 30.4, 117.5, 121.6, and 133.5~nm) with bandwidths of $\sim$1~nm, sampling strong chromospheric and transition-region lines as well as nearby continuum contributions. In particular, channels centered on lines such as He~\textsc{ii} 30.4~nm and H~\textsc{i} Ly-$\alpha$ 121.6~nm include significant continuum background in addition to line emission, allowing these measurements to track variability in optically-thick and semi optically-thick components of the solar spectrum. Although EUVS observations began in 2017 and are expected to continue through 2036, they are not available during the 2010–-2013 period considered here. We therefore generate simulated EUVS measurements by convolving the EVE spectra (0.1~nm resolution) and TIMED-SEE spectra (1~nm resolution) with the EUVS bandpasses\footnote{EUVS bandpasses and responsivities are available from \url{https://data.ngdc.noaa.gov/platforms/solar-space-observing-satellites/goes/goes16/l2/docs/EUVS_responsivity/}.} where EVE spectra are used to simulate the 25.6, 28.4, and 30.4~nm EUVS measurements and TIMED-SEE spectra are used to simulate the 117.5, 121.6, 133.5, and 140.5~nm EUVS measurements. Since the TIMED-SEE spectra are acquired at a 3-minute cadence, the nearest observation to each EVE spectra is used to provide the simulated EUVS measurements at the 60-second cadence. The simulated EUVS measurements are shown in Figure~\ref{fig:Synthetic_EXIS} and are used as inputs to the hybrid model equation from Equation~\ref{equation:hybrid_model} as $I^{proxy}_n$. 

\subsection{$I^{DEM}(\lambda)$ from AIA Data} \label{sec:AIA}

In the hybrid model (Equation~\ref{equation:hybrid_model}), the DEM-based input, $I^{\mathrm{DEM}}(\lambda)$, is derived from coronal emission measured by AIA. Because the model input component is sampled in EVE's 0.1~nm resolution, $I^{\mathrm{DEM}}(\lambda)$ is constructed by first inferring a disk-integrated DEM from the six AIA coronal channels and then forward modeling that DEM into a spectrally resolved irradiance estimate at EVE resolution.

\begin{figure}[h!]
    \centering
    \includegraphics[width=0.43\textwidth, trim={7mm, 4mm, 5mm, 5mm}]{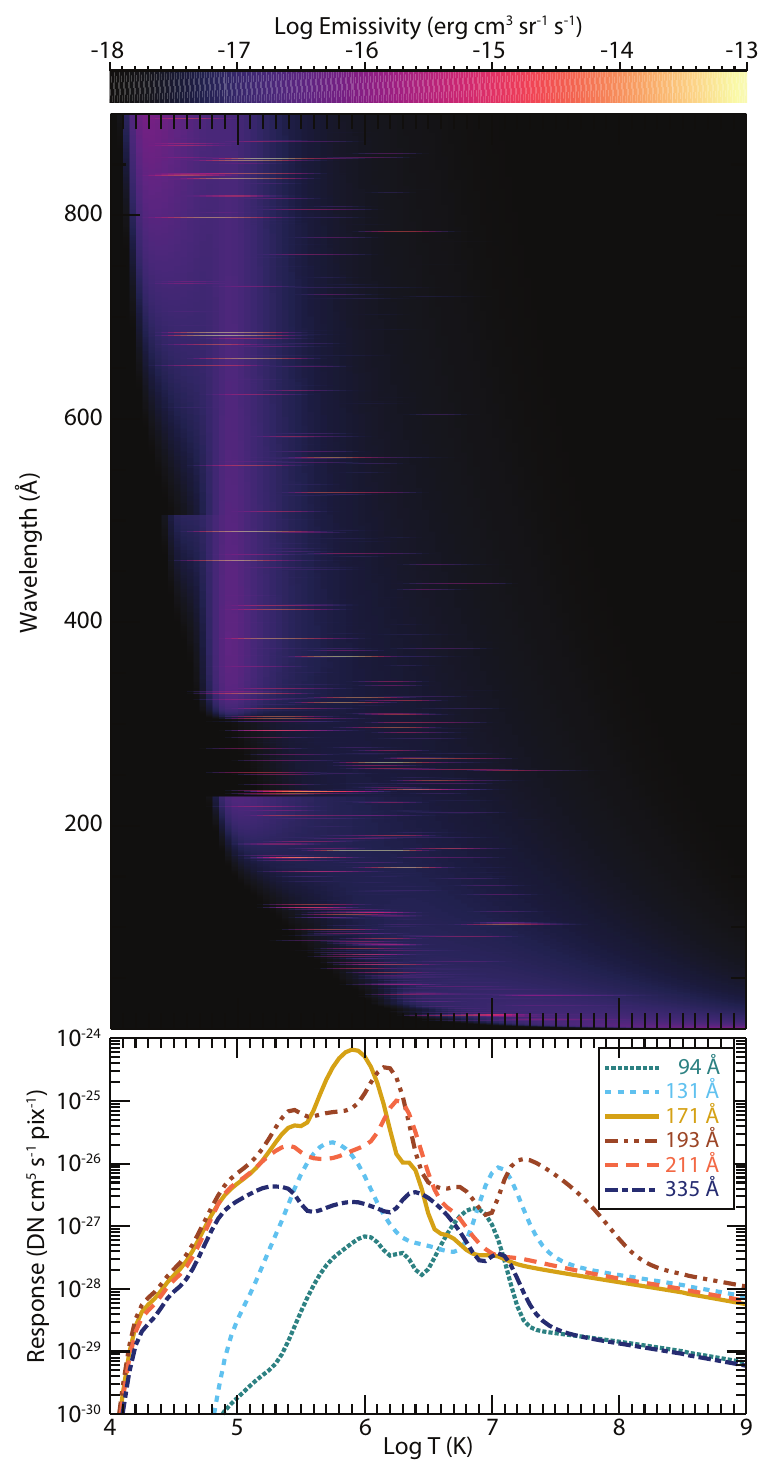}
    \vspace{-3mm}
    \caption{AIA temperature response functions (bottom) compared to modeled plasma emissivity as a function of wavelength and temperature (top) between about 10,000 and $1\times10^9$~K. EUV DEMs provide broad response to a wide range of temperatures and spectral lines.}
    \label{fig:responsivity}
\end{figure}

\begin{figure}[ht!]
    \centering\includegraphics[width=0.47\textwidth, trim={8mm, 10mm, 5mm, 2mm}]{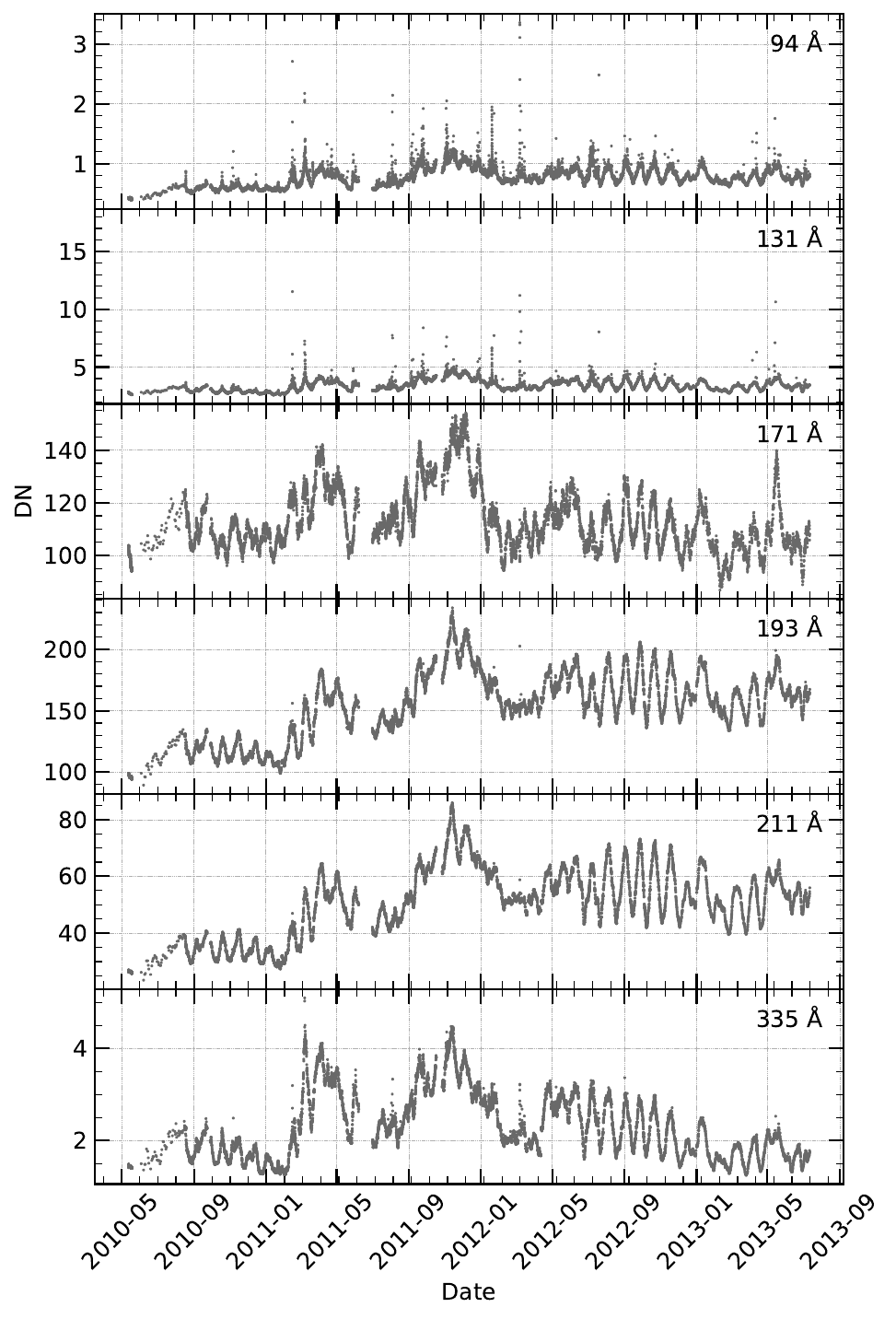}
    \vspace{-3mm}
    \caption{Time series of disk-integrated AIA measurements for the 6 channels measuring optically-thin emission from the corona. }
    \label{fig:Integrated_AIA}
\end{figure}

Degradation-corrected AIA image data are used to calculate the DEMs. AIA acquires seven EUV images with passbands centered on 9.4, 13.1, 17.1, 19.3, 21.1, 30.4, and 33.5~nm at a 12-s cadence with 4096$\times$4096 pixel sampling. With the exception of the 30.4~nm channel, these passbands are dominated by optically-thin coronal emission spanning plasma temperatures from $\sim 10^{4}$~K to $>3\times10^{7}$~K (see the bottom panel of Figure~\ref{fig:responsivity} for the temperature responsivity of the AIA channels, and Table~1 of \citealt{odwyer_sdoaia_2010} for the dominant atomic species and ionization states contributing to each channel). The six coronal channels are therefore used to construct disk-integrated intensity time, shown in Figure~\ref{fig:Integrated_AIA}, which are adopted as the observational constraints for the DEM inversion.
Since DEM inversions utilize the measured emission along the line-of-sight (Equation~\ref{equation:line_of_sight_measurement}), computing DEMs from spatially resolved coronal images and summing the resulting emission is equivalent to first integrating the image intensities over the disk and then performing the DEM inversion on the disk-integrated measurements. In this work, the disk-integrated approach is adopted because the computation time is reduced by approximately the number of AIA pixels (4096$\times$4096), while also providing physically meaningful estimate of the global coronal temperature distribution. Using this disk-integration, the DEM and forward-modeled DEM spectrum can be calculated for an AIA image set in a few seconds rather than hours on a single CPU, enabling the hybrid model approach to be utilized on AIA's 12-s image data cadence. 

\begin{figure}[h!]
    \centering\includegraphics[width=0.48\textwidth, trim={5mm, 10mm, 0mm, 0mm}]{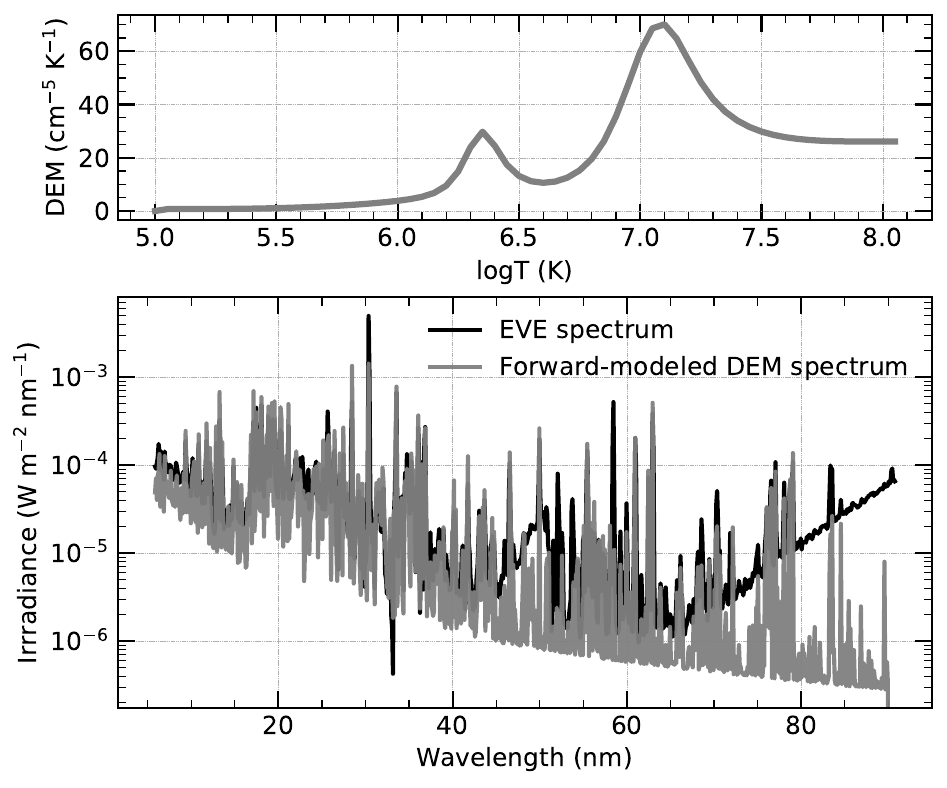}
    \vspace{-3mm}
    \caption{\textit{Top:} DEM generated using spatially-integrated AIA data for the decay phase of an X-class flare on 2012-03-07 from the channels sampling optically-thin EUV emission shown in Figure~\ref{fig:responsivity}. The DEMs were calculated using the Simple Regularized solver from \cite{plowman_fast_2020}. \textit{Bottom:} Forward-modeled spectrum generated from DEM (grey) and corresponding EVE spectrum. }
    \label{fig:DEM}
\end{figure}

For DEM calculation, an emissivity table using CHIANTI v9 and coronal abundances from \citet{schmelz_composition_2012} is computed at 0.01~nm spectral resolution over 0.01--90~nm to ensure sufficient spectral fidelity for forward modeling the DEM spectrum and subsequent resampling to the EVE spectral resolution of 0.1~nm, which is shown in the top panel of Figure~\ref{fig:responsivity}.
The AIA temperature response functions are then calculated using this emissivity table, and are shown in the lower panel of Figure~\ref{fig:responsivity}. Using the AIA temperature response and emissivity table, the DEMs are then inferred from the disk-integrated AIA intensities using the \texttt{simplereg} solver \citep{plowman_fast_2020}. An example DEM derived from disk-integrated AIA data is shown in the top panel of Figure~\ref{fig:DEM}.

The DEM spectrum, $I^{DEM}(\lambda)$, is forward-modeled from the same emissivity table used to calculate the DEMs (shown in top panel of Figure~\ref{fig:responsivity}). This spectrum is then sampled in the EVE spectral resolution of 0.1~nm and is used as the $I^{\mathrm{DEM}}$ input term in Equation~\ref{equation:hybrid_model}. The bottom panel of Figure~\ref{fig:DEM} shows an example forward-modeled $I^{\mathrm{DEM}}$. Optically-thick emission becomes increasingly significant above $\sim$60~nm leading to the increasing divergence between the forward-modeled DEM spectrum and the EVE spectrum in Figure~\ref{fig:DEM}.

\section{Model Training}\label{sec:model}

To assess model performance, the hybrid model is compared against a proxy-only model. 
Both models use the same 13 inputs comprised of full-disk mean intensities from six AIA coronal imaging channels (9.4, 13.1, 17.1, 19.3, 21.1, 33.5~nm) and seven simulated EUVS spectral line channels (25.6, 28.4, 30.4, 117.5, 121.6, 133.5, 140.5~nm). The two models and inputs are summarized below: 
\begin{enumerate}
    \item \textbf{Proxy-only model:} The proxy-only model is used to directly compare the performance of the hybrid method, where the six disk-integrated AIA data and the seven simulated EUVS measurements are the $I^{proxy}_n$ inputs to Equation~\ref{equation:proxy_model} with proxy coefficients $c^{proxy}_n$ and $d$. 
    
    \item \textbf{Hybrid model:} The hybrid model uses forward-modeled spectra from DEMs calculated using the six AIA channel measurements,  $I^{DEM}(\lambda_m)$, and the seven EUVS measurements, $I^{proxy}_n$, as inputs to Equation~\ref{equation:hybrid_model}. The coefficients for the hybrid model are $c^{DEM}_m$, $c^{proxy}_n$, and $d$. Note that the hybrid method and the proxy-only method receive the same input information (the 6 AIA coronal channels and the 7 simulated EUVS proxy measurements). The only difference is that the hybrid model uses the DEM as input rather than the observations from AIA directly, and therefore inherently receives information from the CHIANTI database. 
     
\end{enumerate}

For both models, the proxy  coefficients are determined using multiple linear regression using the \texttt{LinearRegression} module from \texttt{scikit-learn}. Model training and validation use EVE solar spectral irradiance measurements as target outputs over the wavelength range from 5.85~nm to 90~nm, sampled at a fixed 0.1~nm interval. A statistical framework is adopted that is centered on linear regression modeling, in which the train–test split is constructed using dimensionality reduction to manage the spectral complexity of the data, followed by randomized sampling to ensure adequate representation of different solar conditions and events. Section~\ref{sec:data_selection} and ~\ref{sec:data_train_test} in the Appendix describe the data selection and splitting for the training and test datasets. 

In addition to random sampling, we explicitly reserved several well-characterized solar events in their entirety in the test dataset. This was done to evaluate the models' ability to reproduce the full temporal evolution of both flaring and non-flaring conditions. Specifically, we manually selected and set aside the following time periods from the EVE dataset:
\begin{itemize}
  \setlength\itemsep{-0.25em}
    \item A quiet sun interval on 2010/05/16,
    \item An active region passage on 2012/05/16,
    \item An M-class flare event on 2011/02/14, and
    \item An X-class flare on 2012/03/07.
\end{itemize}
Each of these events was extracted with its full set of contiguous time steps, ensuring that their complete rise, peak, and decay phases, or in the case of the quiet sun and active region intervals, their temporal variability, were preserved. These event-based samples were explicitly excluded from the training set to prevent data leakage, thereby allowing for an unbiased assessment of the model’s capacity to generalize to unseen, temporally coherent events.

For each model, the proxy coefficients were derived using the training dataset. Initial evaluations were performed on the training data to ensure convergence and stability of the fits. Final model performance was then independently assessed on the reserved test set using the mean percentage error as the primary evaluation metric. Once validated, each trained model was applied to the entire dataset, including the training and test samples, to generate the modeled spectra over the full timeseries. 

\begin{figure*}[ht!]
    \centering
\includegraphics[width=0.93\textwidth,trim={2mm, 5mm, 3mm, 6mm}]{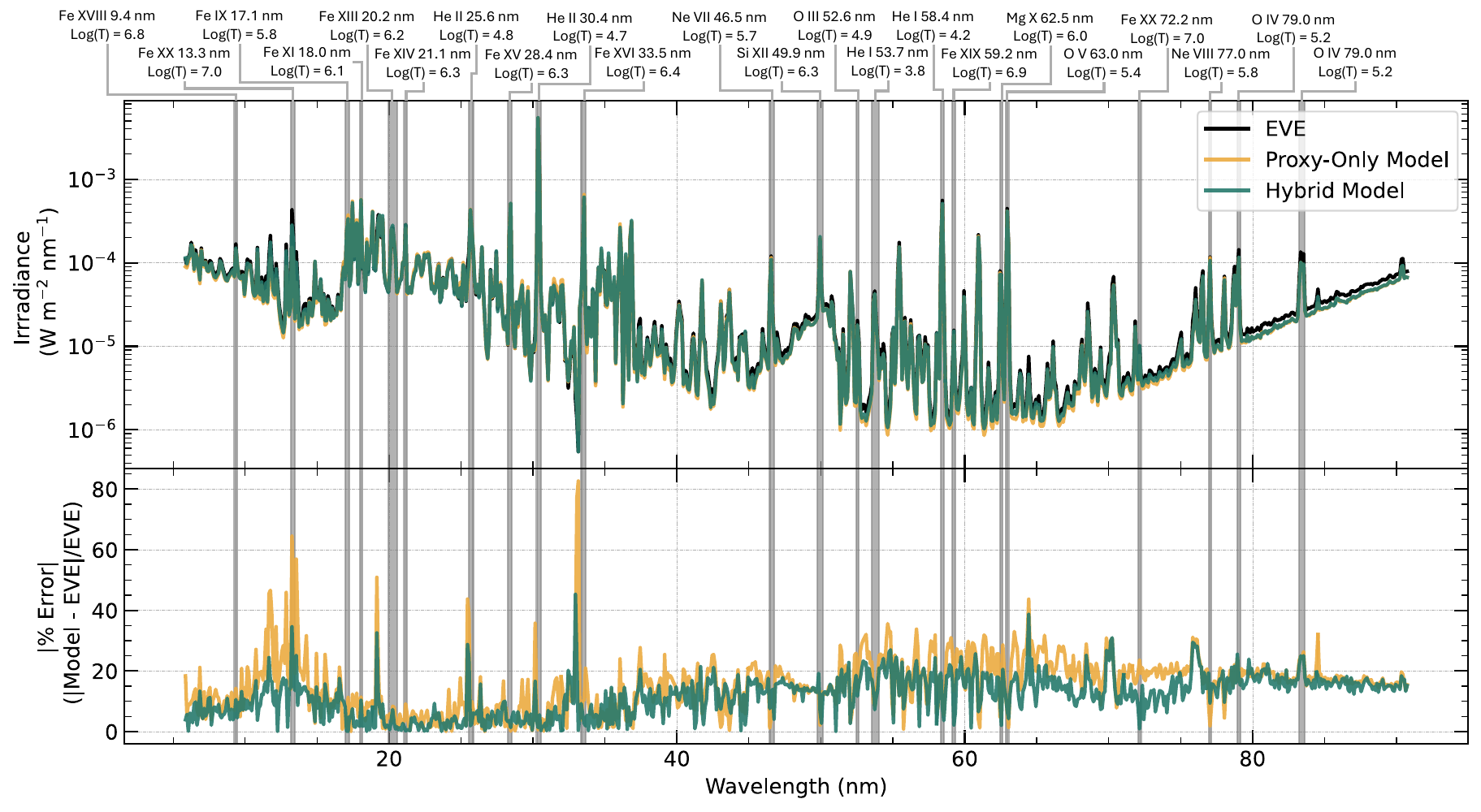}
    \vspace{-3mm}
    \caption{\textit{Top:} 1-hour average spectra generated from hybrid model and proxy-only model for the decay phase of an X-class flare on 2012-03-07. \textit{Bottom:} Fractional error of the modeled spectra relative to EVE. Gray vertical bands represent a subset lines provided in the EVE lines data product and were selected to represent a range of atomic species, ionization states, and formation temperatures.}
    \label{fig:proxy_and_hybrid_spectra}
\end{figure*}

\section{Results} \label{sec:result}

In this section, the modeled spectra from the proxy-only model and the hybrid model are compared against EVE measurements, using the observed spectra as ground truth to calculate percent error of each model. Figure~\ref{fig:proxy_and_hybrid_spectra} shows a 1-hour averaged spectrum, from 5 to 90~nm at 0.1~nm resolution, generated by the proxy-only and hybrid models for the decay phase of an X-class flare on 2012-03-07, together with the corresponding EVE spectrum. In addition, model performance is assessed over 22 spectral lines annotated in the top panel of Figure~\ref{fig:proxy_and_hybrid_spectra}, chosen to provide representative coverage of different atomic species, ionization states, and formation temperatures; the highlighted lines are a subset of those provided in the EVE Level 2B lines data product. The lower panel presents the absolute percentage error as a function of wavelength for both models relative to the EVE measurements. It should be noted that absolute percentage errors are particularly sensitive to deviations in spectral regions where the signal level is low; consequently, larger relative errors generally occur in troughs of the spectrum.

\begin{figure*}[t!]
    \centering
\includegraphics[width=0.95\textwidth]{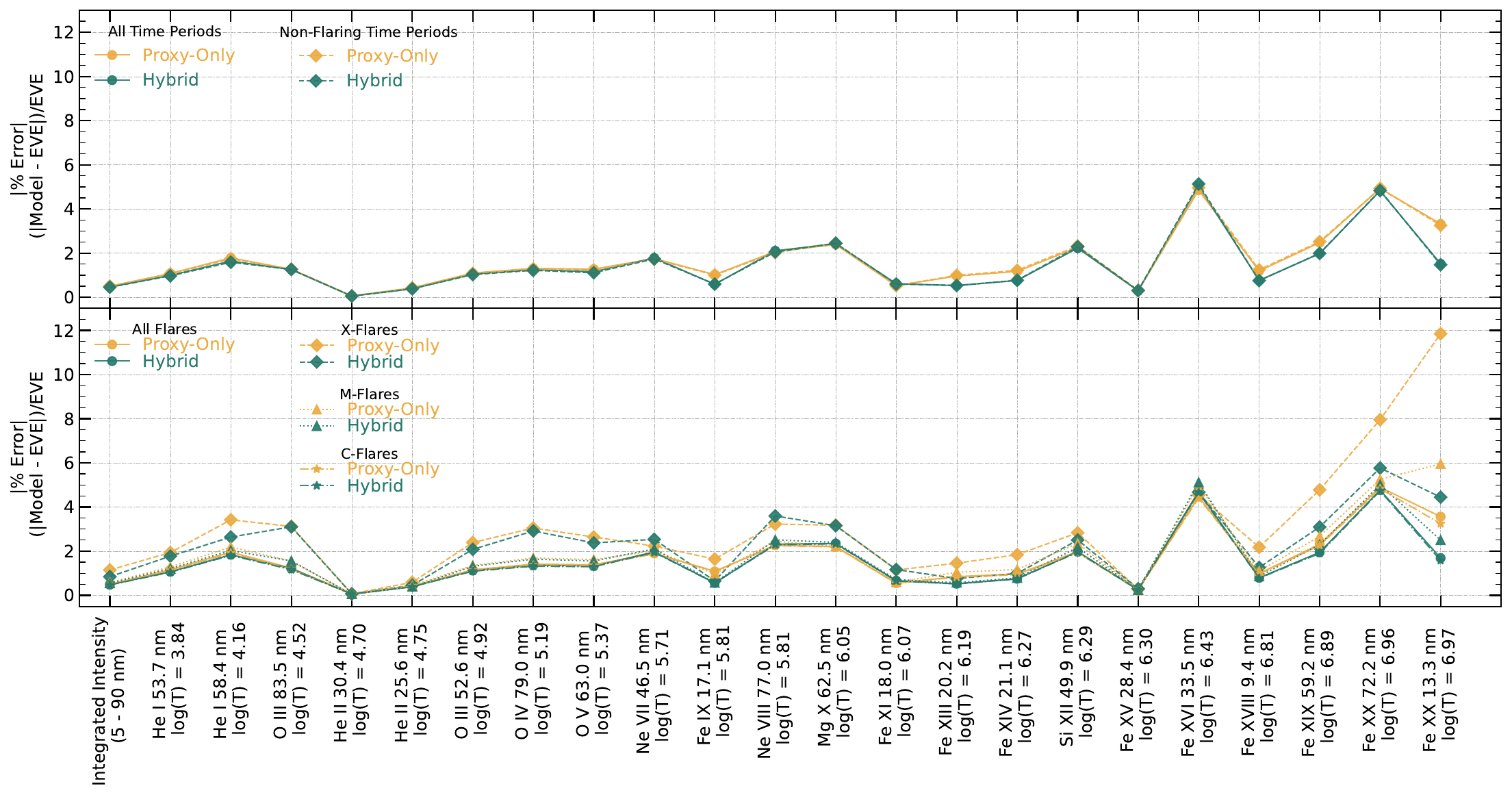}
    \vspace{-3mm}
    \caption{Average fractional error of the hybrid and proxy-only models relative to EVE measurements for the 5 -- 90~nm integrated intensity and line-integrated irradiances. The fractional errors are averaged over time periods of selected solar conditions \textit{Top:} all time periods and non-flaring time periods, \textit{Bottom:} all flares, X-class, M-class, and C-class flares.}
    \label{fig:aggregate_errors}
\end{figure*}

\subsection{Time-Averaged Model Performance Across Solar Conditions}

To investigate temperature sensitivity across different heights of the solar atmosphere, both models are evaluated using the same 22 diagnostically important emission lines shown in Figure~\ref{fig:proxy_and_hybrid_spectra}, with known formation temperatures spanning $\log T \approx 3.8$--7.0. To examine how model performance depends on solar activity, the full time series from 2010 to 2013 is partitioned into six categories: all-time, non-flaring, flaring, X-flares, M-flares, and C-flares, using the V1 Level~2 GOES XRS flare summary product \texttt{sci\_xrsf-l2-flsum} (Machol et al. 2026, submitted). This product provides the flare class as well as the flare start, peak, and end times, which are used to define the flaring intervals for each category based on the start-to-end time windows (see Section~\ref{sec:data_selection} for additional details). Within each solar activity category, the model errors are temporally integrated for each of the 22 spectral lines and for the spectrally-integrated intensity from 5--90~nm. 

Figure~\ref{fig:aggregate_errors} summarizes the mean absolute percentage errors of the proxy-only and hybrid models for the solar activity categories across the the spectrally-integrated intensity and the 22 spectral lines ordered by formation temperature. For all-time and non-flaring periods (top panel), the two models exhibit nearly identical performance across the full set of lines, with a moderate improvement shown by the hybrid model in the high-temperature regime (log $T \geq 6.81$). In contrast, substantial differences in absolute percentage error emerge during flaring conditions (bottom panel), where the hybrid model consistently outperforms the proxy-only model, with the largest improvements occurring during X-class flares. In this regime, the proxy-only errors increase systematically toward higher formation temperatures, whereas the hybrid model maintains relatively uniform error levels across the temperature range. 

This behavior is most evident for the hottest coronal lines: for example, during X-class flares the error in Fe~\textsc{xx} (log $T \approx 6.97$) is reduced from $\sim$12\% in the proxy-only model to $\sim$4.5\% in the hybrid model, for Fe~\textsc{xix} (log $T \approx 6.89$) from $\sim$5\% to $\sim$3\%, and for Fe~\textsc{xviii} (log $T \approx 6.81$) from $\sim$2\% to $\sim$1.2\%. More generally, the separation between the two models increases with flare class, with the largest improvement observed for X-class events and a more moderate but still systematic improvement for M- and C-class flares. At lower formation temperatures, including lines such as He~\textsc{i} (58.4~nm), O~\textsc{v} (63.0~nm), and O~\textsc{iii} (83.5~nm), the difference between the two models is less pronounced. For these cooler lines representing chromospheric and transition region emission, the proxy-only and hybrid models achieve similar results, reflecting the fact that in this regime the hybrid model is dominated by the proxy component because the DEMs cannot constrain the optically thick plasma. Overall, the figure highlights that the principal gains of the hybrid approach are concentrated in flaring conditions, particularly for high-temperature coronal diagnostic lines.

The reduced errors exhibited by the hybrid model at higher formation temperatures reflect the benefit of explicitly encoding the coronal thermal structure through DEM-based forward modeling of optically thin line emissions. In contrast, the proxy-only model does not incorporate a direct, temperature-dependent physical description of the emitting plasma and therefore shows systematically increasing errors toward hotter lines, particularly during flaring conditions. The results presented here demonstrate that incorporating physically motivated thermal information substantially improves spectrally-resolved irradiance modeling, especially during flare-dominated intervals when hot coronal plasma ($\log T \gtrsim 6.7$) dominates the emission.

\subsection{Time-Resolved Model Performance}

\begin{figure*}[ht!]
    \centering
\includegraphics[width=0.92\textwidth]{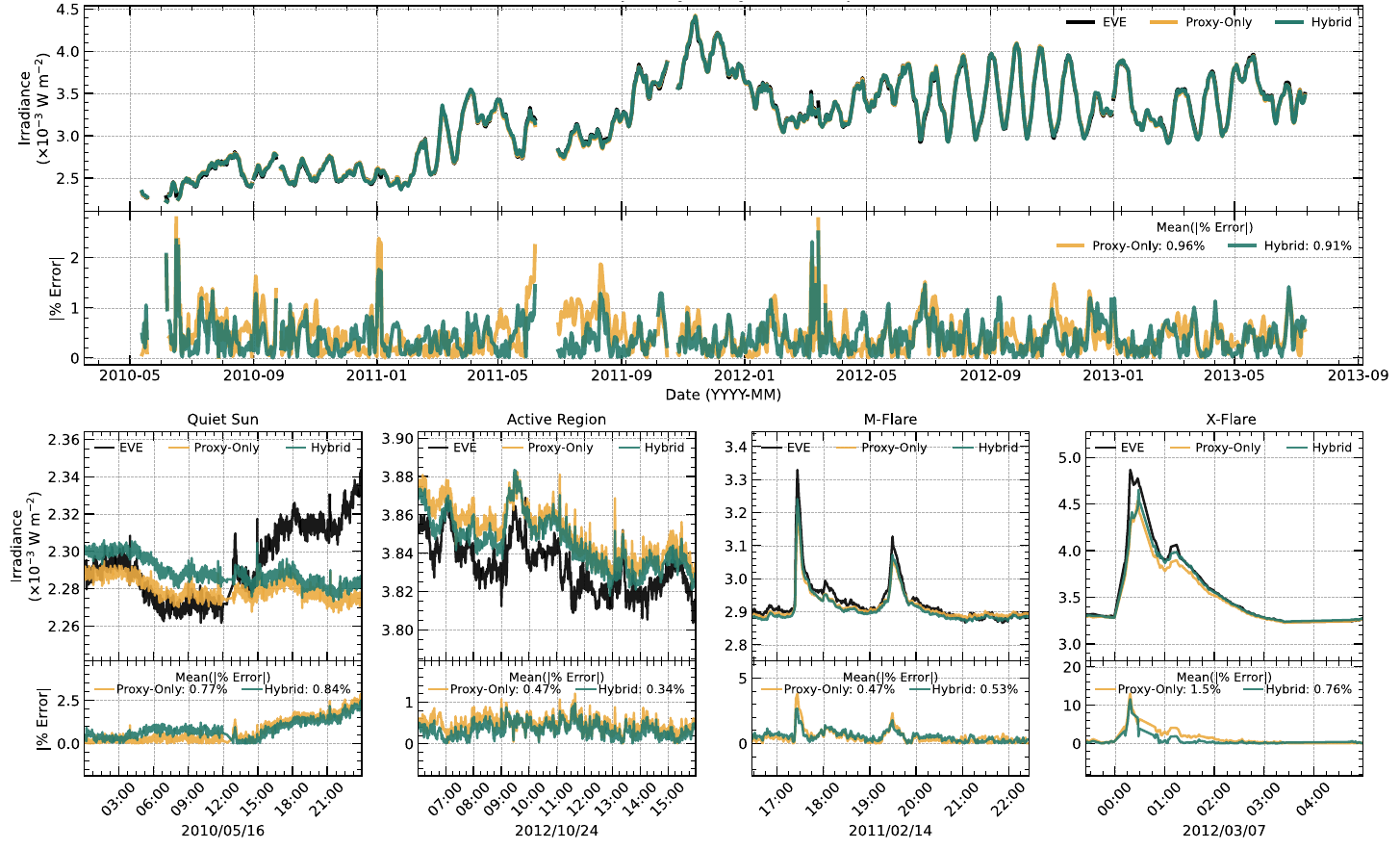}
    \vspace{-3mm}
    \caption{\textit{Top:} Daily averaged predicted integrated intensity (5 -- 90~nm) for the hybrid and proxy-only models and EVE measurements (upper panel) and the corresponding percent error (lower panel).  \textit{Bottom:} Predicted integrated intensity for the hybrid and proxy-only models and EVE measurements for four periods of differing activity states -- quiet sun, active region, M-flare, X-flare -- (upper panels) and the corresponding percent errors relative to EVE (lower panels). Percent error legends provide the  average percent error over the graphed time range. Note that the timeseries in the bottom figures were excluded from model training.}\label{fig:integrated_intensity_results}
\end{figure*}

To evaluate the model performance across the full spectral range, the integrated solar EUV irradiance from 5 to 90~nm is computed for both the hybrid and proxy-only models and is compared against EVE reference spectra. Figure~\ref{fig:integrated_intensity_results} illustrates this comparison for the daily-averaged time series from 2010~--~2013 (top) and for the reserved solar conditions (bottom): quiet sun, active region, M-class, and X-class flaring intervals, which were withheld from the training data as described in Section~\ref{sec:data_train_test}. Over the entire dataset, the hybrid model achieves a mean error of 0.91\%, slightly outperforming the 0.96\% error from the proxy-only model. The two models perform similarly for the quiet sun, active region, and M-flare time periods, while the hybrid model shows a clear improvement over the proxy-only model in reconstructing the X-flare irradiance (0.76\% vs. 1.5\%). These findings underscore the hybrid model's ability to generalize across a broad range of irradiance conditions, particularly when capturing impulsive, thermally complex emission.

\begin{figure*}[h!]
    \centering
\includegraphics[width=0.90\textwidth]{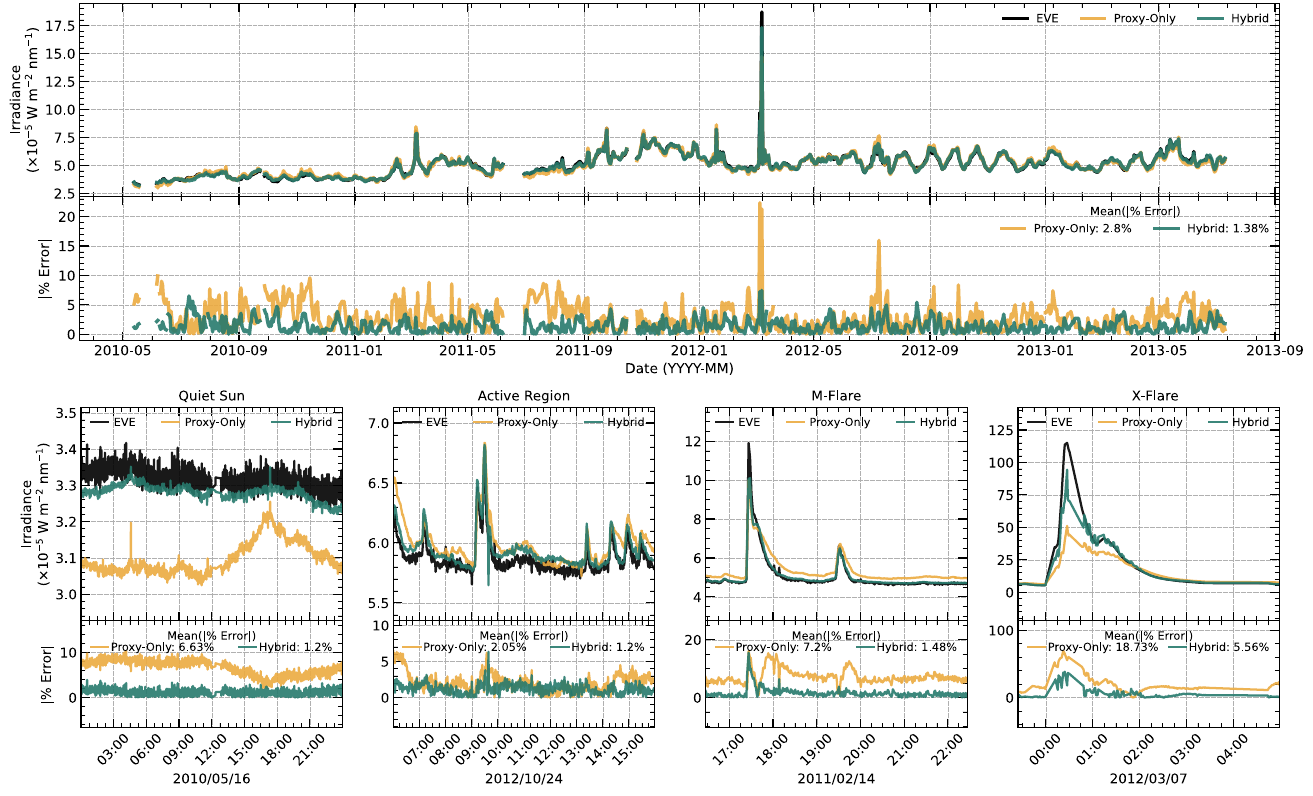}
    \vspace{-3mm}
    \caption{Same as Figure~\ref{fig:integrated_intensity_results} but for the Fe~\textsc{xx}~(13.3~nm) spectral line (logT~$=$~6.97).} ~\label{fig:Fe13.3_results}
\end{figure*}

\begin{figure*}[h!]
    \centering
\includegraphics[width=0.90\textwidth]{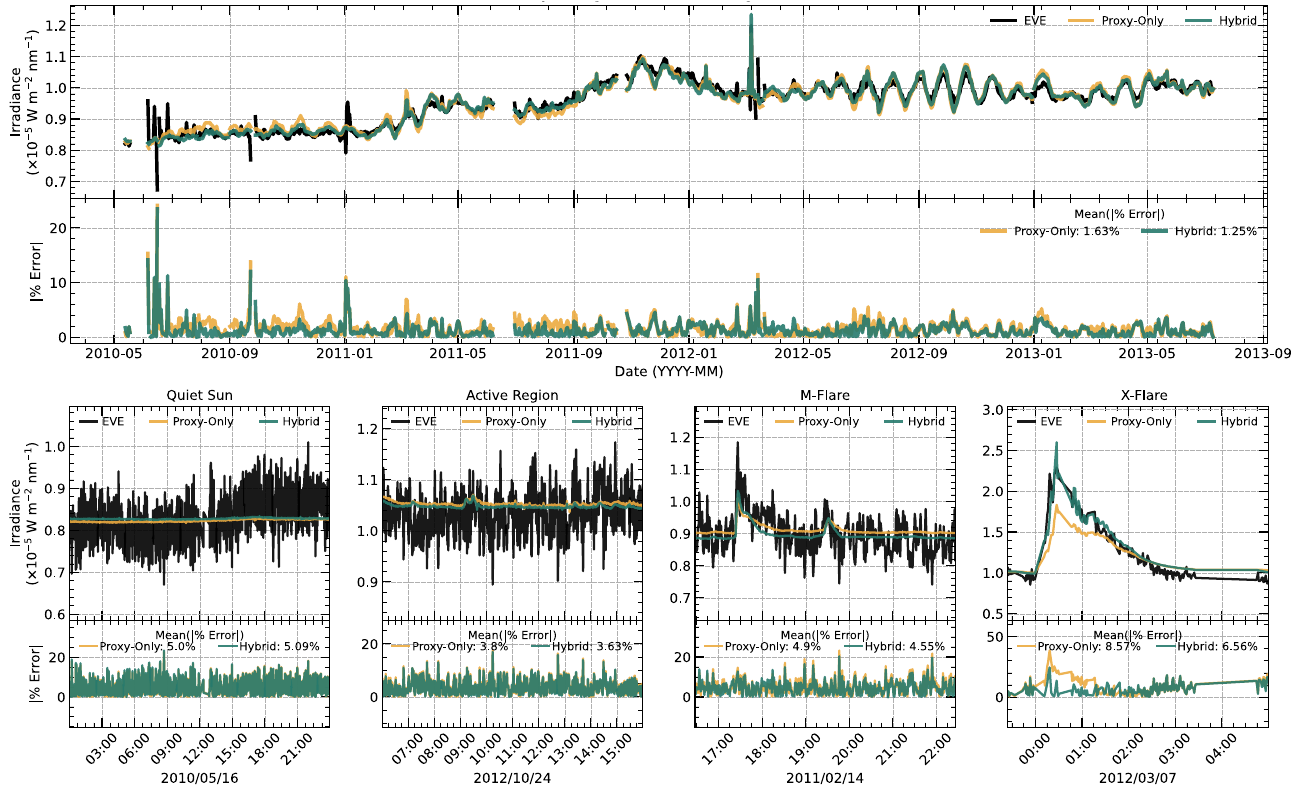}
    \vspace{-3mm}
    \caption{Same as Figure~\ref{fig:integrated_intensity_results} but for the Fe~\textsc{xx}~(72.2~nm) spectral line (logT~$=$~6.96).} ~\label{fig:Fe72.2_results}
\end{figure*}

\begin{figure*}[h!]
    \centering
\includegraphics[width=0.90\textwidth]{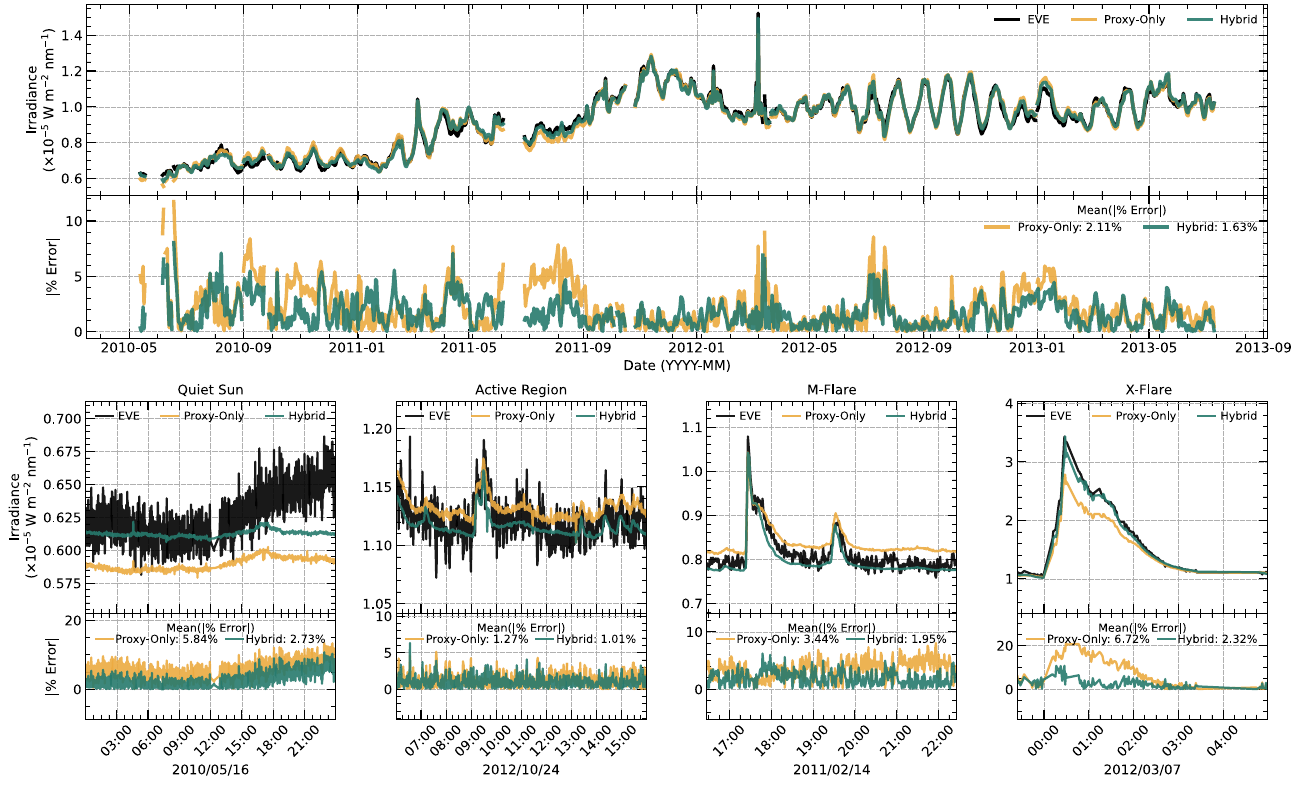}
    \vspace{-3mm}
    \caption{Same as Figure~\ref{fig:integrated_intensity_results} but for the Fe~\textsc{xix}~(59.2~nm) spectral line (logT~$=$~6.89).} ~\label{fig:Fe59.2_results}
\end{figure*}

\begin{figure*}[h!]
    \centering
\includegraphics[width=0.90\textwidth]{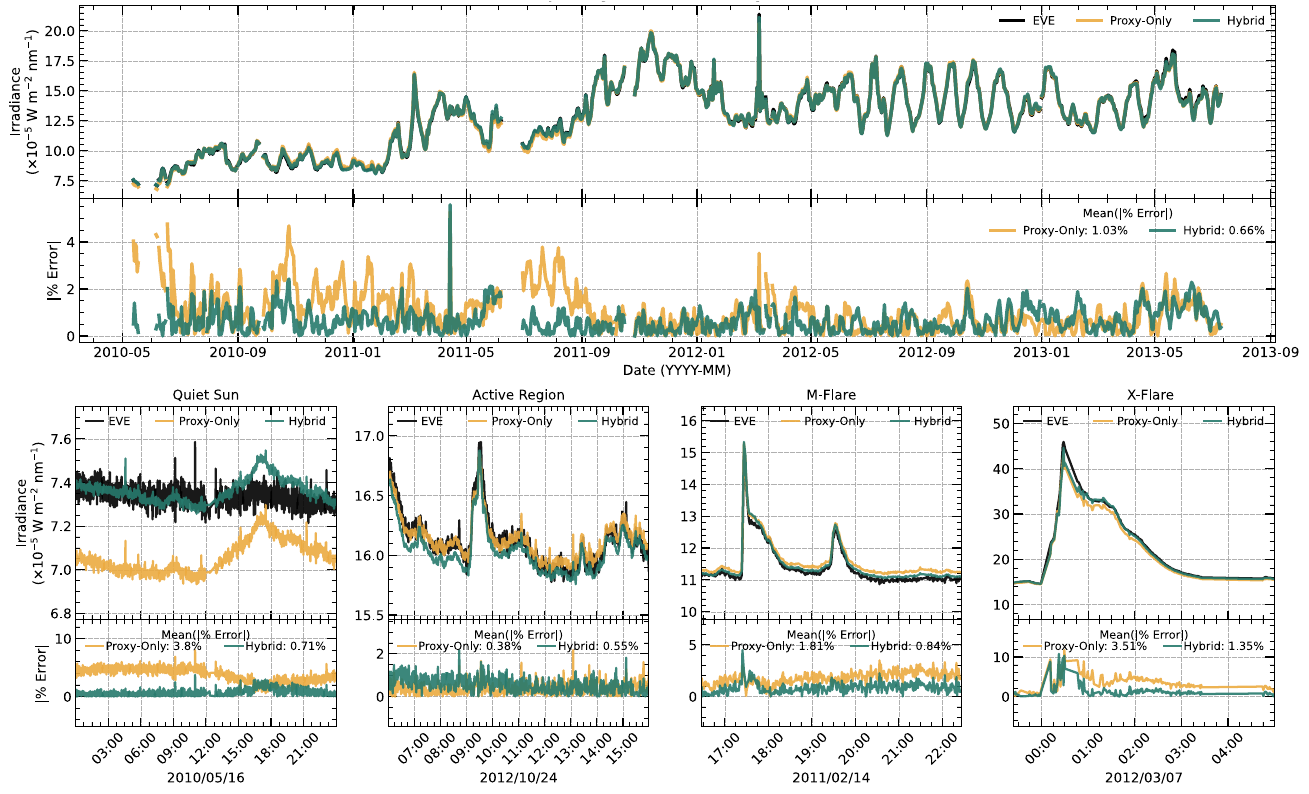}
    \vspace{-3mm}
    \caption{Same as Figure~\ref{fig:integrated_intensity_results} but for the Fe~\textsc{xviii}~(9.4~nm) spectral line (logT~$=$~6.81).} ~\label{fig:Fe9.4_results}
\end{figure*}

\begin{figure*}[h!]
    \centering
\includegraphics[width=0.90\textwidth]{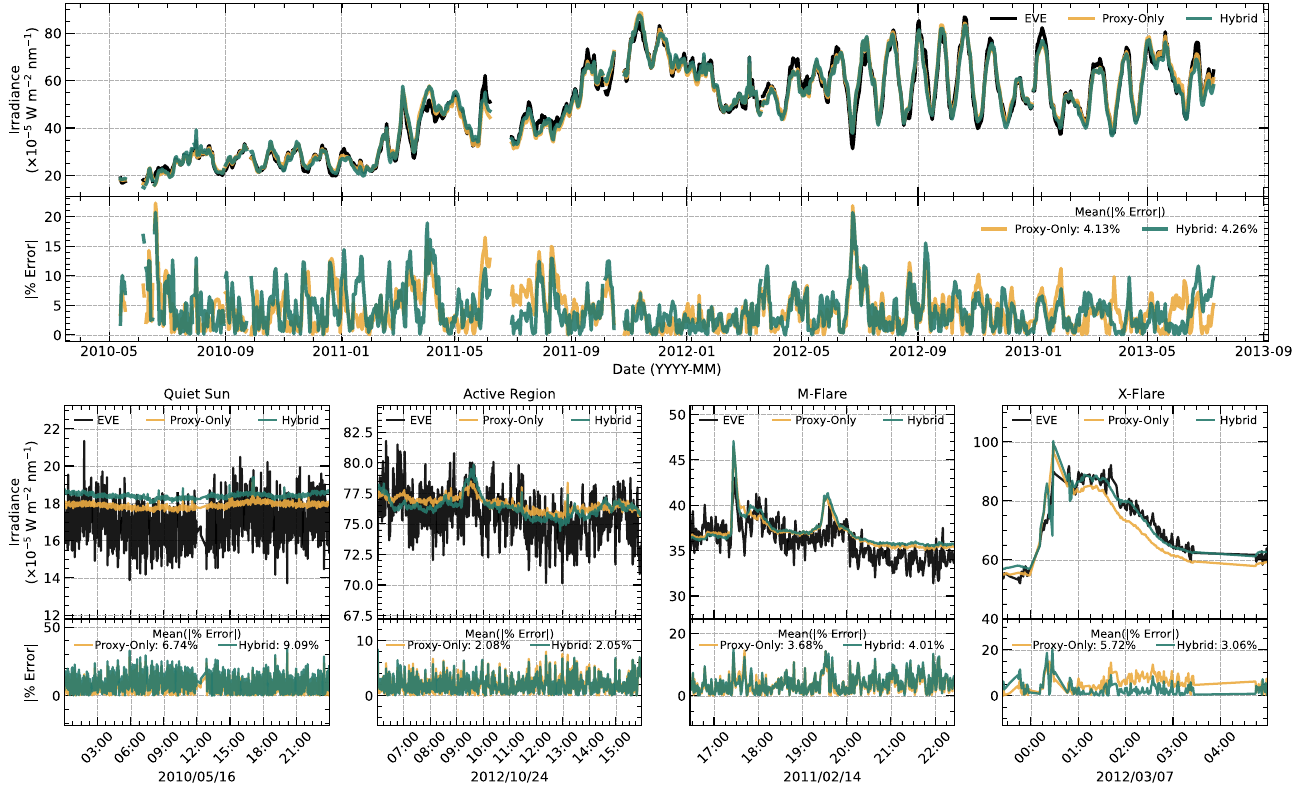}
    \vspace{-3mm}
    \caption{Same as Figure~\ref{fig:integrated_intensity_results} but for the Fe~\textsc{xvi}~(33.5~nm) spectral line (logT~$=$~6.43).} ~\label{fig:Fe33.5_results}
\end{figure*}

\begin{figure*}[h!]
    \centering
\includegraphics[width=0.90\textwidth]{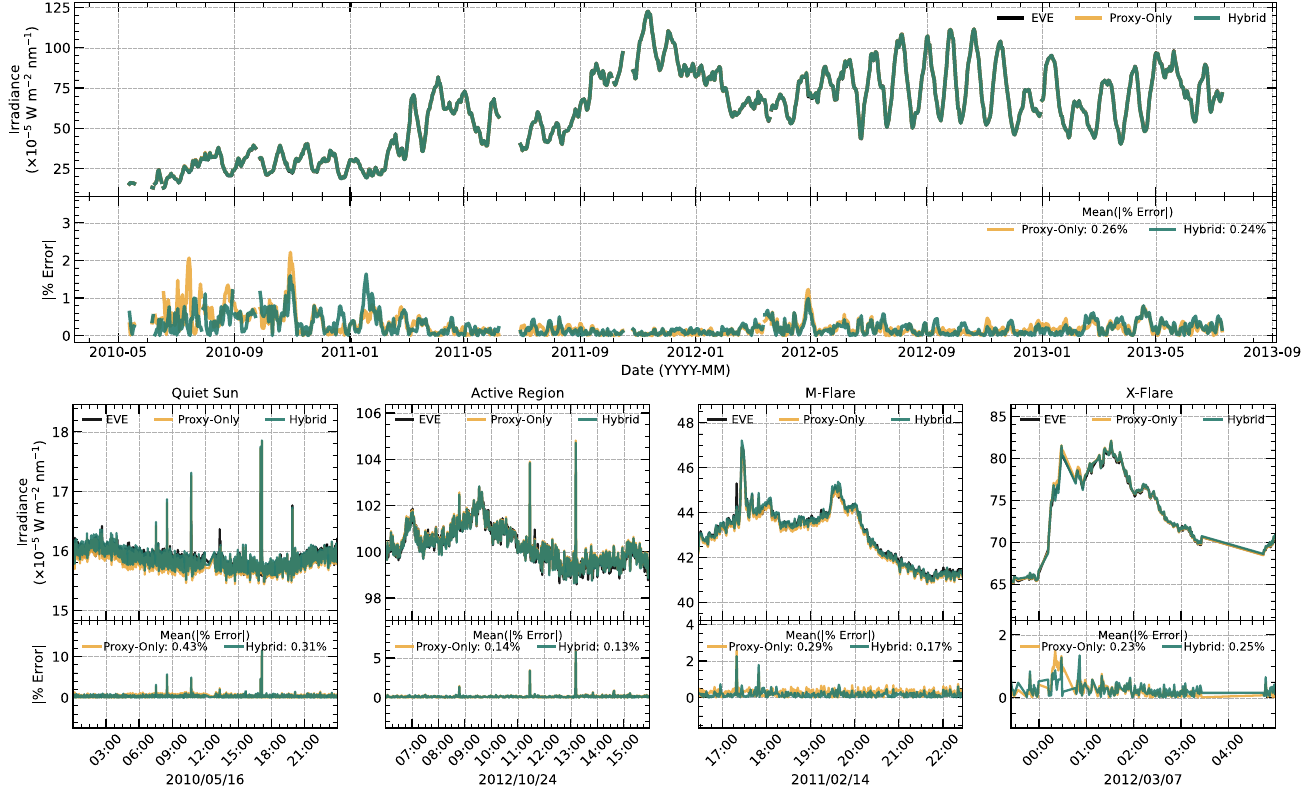}
    \vspace{-3mm}
    \caption{Same as Figure~\ref{fig:integrated_intensity_results} but for the Fe~\textsc{xv}~(28.4~nm) spectral line (logT~$=$~6.30).} ~\label{fig:Fe28.4_results}
\end{figure*}

Figures~\ref{fig:Fe13.3_results}, \ref{fig:Fe72.2_results}, \ref{fig:Fe59.2_results}, \ref{fig:Fe9.4_results}, \ref{fig:Fe33.5_results}, and \ref{fig:Fe28.4_results} show the same time series comparisons as Figure~\ref{fig:integrated_intensity_results}, but for the spectral lines with $\log T >6.3$ sorted by formation temperature (high to low). These hotter lines are selected to be discussed in this section because the results in Figure~\ref{fig:aggregate_errors} show the largest improvement in performance in this temperature regime. The figures for four additional cooler spectral lines are provided in Appendix~\ref{sec:cool_line_performance}, while the time-averaged performance of all 22 spectral lines is provided in Figure~\ref{fig:aggregate_errors}.


For Fe~\textsc{xx} (13.3~nm), shown in Figure~\ref{fig:Fe13.3_results}, the hybrid model demonstrates a substantial improvement over the proxy-only model across all tested solar conditions. In the top panel, the hybrid model spectra follows the EVE reference time series more closely, particularly during periods of rapid variability and enhanced emission. This improvement is reflected quantitatively in the lower panels, where the hybrid model exhibits more than three-fold reduction in the mean absolute percent error compared to the proxy-only model for the quiet Sun, M-flare, and X-flare test cases, and approximately a two-fold reduction for the active region. In particular, for the M- and X-class flare intervals, the mean absolute percentage error decreases from 7.2\% to 1.48\% and from 18.73\% to 5.56\%, respectively. 

For Fe~\textsc{xx} (72.2~nm), shown in Figure~\ref{fig:Fe72.2_results}, both models capture the overall temporal evolution of the irradiance across the full time series and for the selected solar conditions. In the top panels, the hybrid model follows the EVE ground truth more closely during periods of enhanced variability, while the proxy-only model tends to underestimate the peak emission. This behavior is reflected in the lower panels, where the hybrid model achieves a systematically lower mean absolute percent error, notably for the X-class flare case, where the error is reduced from 8.57\% to 6.56\%. More modest improvements are also seen for the M-flare cases, while the two models otherwise exhibit comparable performance during the quiet Sun and active region intervals. Overall, the gains for Fe~\textsc{xx} (72.2~nm) are present but smaller than those seen for the shorter-wavelength, hotter coronal lines discussed below (Figures~\ref{fig:Fe59.2_results} and \ref{fig:Fe9.4_results}).

Figures~\ref{fig:Fe59.2_results} and \ref{fig:Fe9.4_results} show the corresponding comparisons for Fe~\textsc{xix} (59.2~nm) and Fe~\textsc{xviii} (9.4~nm), respectively. For both lines, the hybrid model consistently provides a closer match to the EVE reference time series, particularly during flaring intervals, as seen in the top panels. These improvements are quantified in the lower panels by a reduction in the mean absolute percent error across all test cases. For Fe~\textsc{xix} (59.2~nm), the mean absolute percent error decreases from 6.72\% to 2.32\% for the X-flare case and from 3.44\% to 1.95\% for the M-flare case, while smaller, but consistent improvements are also obtained for the quiet Sun and active region intervals. Similarly, for Fe~\textsc{xviii} (9.4~nm), the hybrid model reduces the mean absolute percent error from 3.51\% to 1.35\% for the X-flare case and from 1.81\% to 0.84\% for the M-flare case, with corresponding improvements under quiet sun conditions. Together, these improvements are consistent with the reduction reported in Figure~\ref{fig:aggregate_errors}, which summarizes performance over the entire dataset, including both training and test intervals.

Figures~\ref{fig:Fe33.5_results} and \ref{fig:Fe28.4_results} present the time-resolved comparisons for Fe~\textsc{xvi} (33.5~nm; $\log T = 6.43$) and Fe~\textsc{xv} (28.4~nm; $\log T = 6.30$), probing relatively cooler coronal plasma than the lines discussed above. For both lines, the two models closely track the EVE reference time series across the full dataset, as shown in the top panels, indicating that the dominant temporal variability is well captured in both cases. Consistent with this visual agreement, the lower panels show only small differences in mean absolute percent error between the two approaches. For Fe~\textsc{xvi} (33.5~nm), the hybrid model yields a slightly higher error during the quiet Sun interval (9.09\% vs.\ 6.74\%), while achieving similar performance for the active region and M-flare cases and a clear improvement for the X-flare case (5.72\% to 3.06\%). For Fe~\textsc{xv} (28.4~nm), both models perform at a very high level across all activity regimes, with mean absolute percent errors below 0.5\% in most cases and only marginal differences between the two approaches. The small magnitude and relative stability of the mean absolute percent error in this regime are consistent with the fact that emission at these temperatures is well sampled by the proxy inputs used by both models (33.5~nm sampled by AIA and 28.4~nm sampled by the simulated EUVS data), such that the irradiance in this regime is already well constrained by the available observables.

Across most spectral lines and time intervals, the hybrid model reproduces the temporal structure and amplitude of the EVE-measured irradiance more accurately than the proxy-only model. Importantly, the observed accuracy improvements from the hybrid model hold regardless of whether the spectral line in question is derived from an AIA-observed band or simulated EUVS input data. While some lines, such as those near 13.3~nm and 33.5~nm, are directly constrained by the AIA channels used in the DEM inversions, others such as the optically thick He~\textsc{ii}~30.4~nm line (Figure~\ref{fig:He30.4_results}) are informed by the simulated EUVS measurements. Other lines, such as Ne~\textsc{vii}~46.5~nm (see Figure~\ref{fig:Ne46.5_results}), are not directly measured by AIA or EUVS and are therefore informed by the proxy relationships.

In many cases, particularly for the quiet sun test case, the reconstructed irradiance for both models shows noticeably less high-frequency variability than the EVE measurements, which is an expected outcome for regression-based models. The short-timescale fluctuations in EVE, arising from photon statistics, detector readout, and intrinsic solar variability, have no linear correlation with the proxies and are therefore naturally suppressed by the regression. As a result, the model output approximates the conditional mean of the irradiance given the inputs, yielding a smooth estimate of the underlying solar EUV variability rather than the instantaneous noise present in the raw measurements.

Despite the overall performance advantage of the hybrid model, there are instances where the proxy-only model outperforms the hybrid approach, particularly for lower-temperature lines in the quiet sun test case. For example, in lines such as Fe~\textsc{ix}~(17.1~nm), Fe~\textsc{xvi}~(33.5~nm), and Si~\textsc{xii}~(49.9~nm) in Figures~\ref{fig:Fe17.1_results}, \ref{fig:Fe33.5_results}, \& \ref{fig:Si49.9_results} respectively, the hybrid model exhibits slightly elevated error compared to the proxy-only model in the quiet sun test case. These discrepancies may not necessarily indicate a failure of the model to accommodate the underlying physics, but may instead arise from limitations in the DEM reconstruction process when the signal-to-noise ratio is low, as is often the case during quiet sun periods. Additionally, unresolved measurement noise in the EVE data may disproportionately affect these lines, especially when the irradiance is low. Since both models are trained on temporally filtered and flare-stratified datasets, the hybrid model may under-fit features that are poorly constrained during quiet periods. Future work will investigate the noise floor and other potential sources of systematic error specific to quiet sun conditions and incorporate these effects into the modeling framework to improve reconstruction accuracy across all activity levels.

Relative to the pre-flare error, increased error of the hybrid model is observed in almost all lines during rapid transitions, particularly during the rise phase and, in some cases, the decay phase of the X-class flare test case. This suggests that limitations in temperature coverage, cadence, or accuracy of the model input data may affect the DEM’s effectiveness for reconstructing spectral lines at all temperatures, and that even the hybrid model is not fully equipped to encapsulate the temporal and thermal complexities associated with high-temperature optically-thin lines and low-temperature semi optically-thick lines of the solar atmosphere concurrently. Since the hybrid model performs more reliably for C- and M-class flares than for X-class flares (see Figure~\ref{fig:aggregate_errors}), this limitation is likely due in part to saturation of the AIA detector during flares $\gtrapprox$ X1 \citep{Guastavino2019} that limits the accuracy of DEM reconstructions. Despite these limitations, the hybrid model outperforms the proxy-only model in nearly all spectral lines for the both X-class flare test case (Figures~\ref{fig:Fe13.3_results} -- \ref{fig:He30.4_results}) and in the integrated performance over all X-class flares (Figure~\ref{fig:aggregate_errors}).

\section{Conclusions} \label{sec:conclusions}

DEM-driven spectral reconstructions provide a physically motivated approach to estimate solar EUV spectral irradiance, particularly in regimes dominated by optically-thin coronal plasma, such as during solar flares. At the same time, proxy-based approaches remain essential for capturing emission from optically-thick regions of the lower solar atmosphere, where proxies from DEM-derived spectra are not applicable and empirical relationships often have the highest fidelity. The results presented here demonstrate that neither approach alone is sufficient to achieve robust spectral reconstructions across all solar conditions. Instead, combining DEM-based coronal information with proxy-based constraints yields systematically improved performance, especially for high-temperature, flare-dominated regimes, while preserving accuracy at lower temperatures where proxy relationships remain effective. Our hybrid framework therefore provides a more complete and reliable representation of the solar EUV spectrum and shows potential to provide useful estimates of short-wavelength solar spectral irradiance without the need for dedicated instrumentation.

The model is not without limitations, however. For example, it is important to note that AIA observations are often saturated during large flares, depending on exposure settings and flare duration \citep[see discussion in][]{Guastavino2019}. While saturation can distort spatial information and obscure detailed morphological analysis, its impact on the hybrid model approach is mitigated to some extent by using disk-integrated intensity values rather than localized pixel-level structures. If the saturation redistributes flux across neighboring pixels (i.e., blooming or smearing), the total integrated intensity may remain a reasonable proxy for the true emission measure. However, in cases where saturation leads to intensity clipping and a net loss of signal from high-emission regions, this can result in an underestimation of the thermal energy content in the DEM calculation. Consequently, the spectra generated from such DEMs would misrepresent the true spectral output during flare peaks. In such case, the degraded DEM reconstructions under saturation conditions could propagate errors into the forward-modeled spectra and thereby reduce the accuracy of the predicted irradiance in the hybrid model. 

Addressing this limitation for AIA data may require saturation-aware pre-processing or correction algorithms, including flare-sensitive proxies that could fill gaps in DEM sensitivity for the biggest events, in future iterations of this modeling framework. An alternative approach would be to use high-dynamic-range images, like those from the GOES-R Solar UltraViolet Imager \citep[SUVI;][see section 3.2.1]{darnel_goes-r_2022}, which are effectively immune to the limitations of saturation -- although such a change would come at the expense of decreased cadence due to the way in which SUVI's observing operations are implemented.

From an operational space weather perspective, this hybrid framework is compelling because it can be implemented using existing real-time solar observations due to its fast execution time from calculating DEMs on the disk-integrated EUV image data. In principle, the combination of coronal imaging from SUVI and spectral irradiance measurements from the EXIS EUVS instruments aboard the GOES-R series provides the necessary inputs to drive such a model in near real time. This capability opens a path toward routine, physics-informed reconstructions of the solar EUV spectrum for both scientific analysis and space weather applications, without reliance on continuous high-resolution spectrographs. Such an adaptation will be  the focus of future work on this project. 

In fact, DEM techniques have already been demonstrated using SUVI data \citep[see][Figure~19]{darnel_goes-r_2022}, although additional challenges arise relative to AIA-based reconstructions due to channel cross-contamination in some SUVI passbands. Building on these developments, future work will combine SUVI-based DEMs with EUVS narrowband spectral measurements to extend the hybrid framework presented here, which was demonstrated using AIA and EVE (with EVE integrated to simulate EUVS measurements). This step is necessary because SUVI and EUVS together constitute a long-lived, operational observing system on the GOES-R series, providing continuous real-time measurements of the solar EUV proxies through SWPC until at least 2036. This model is currently under development and requires bootstrapping the model calibration here into the EXIS and SUVI era. 

In addition to the hybrid model presented in this work, we explored alternative model configurations to assess the robustness of different approaches for EUV spectra modeling. One variant involved replacing the forward-modeled spectra with the differential emission measure (DEM) itself as an input. However, this approach yielded worse performance relative to the full hybrid method, likely due to the loss of wavelength-dependent information for individual ions encoded in the forward modeling step. We also tested a simple feedforward neural network (NN) framework, trained separately on both forward-modeled spectra and proxy-only inputs. While the NN model achieved comparable performance to the hybrid approach in some configurations, it demonstrated a higher tendency to overfit, particularly during strong flares, and offered less interpretability. Based on these comparisons, we conclude that the hybrid model described in this paper provides the most consistent performance across different solar activity levels while maintaining physical transparency.

Interestingly, our analysis indicates that proxy-based models often outperform DEM-based approaches during quiet sun conditions, likely due to the increased uncertainty in DEM inversion during quiescent solar activities. In contrast, the hybrid model consistently outperforms the proxy-only model during flare events, particularly for high-temperature spectral lines. These complementary characteristics suggest that different modeling strategies may be better suited to different regions of parameter space. This opens the door to future ensemble-based modeling frameworks, in which multiple models can be trained independently and selected dynamically based on the input features. While the design and implementation of such an adaptive machine learning model is beyond the scope of the present study, we plan to pursue this approach in future studies.

\vspace{5mm}
\noindent
This work was supported by NASA grant award number 80NSSC20K1363. 
\vspace{5mm}

%


\noindent \textit{Data Acknowledgments:} \\
SDO EVE Level 2B spectra and lines data products: \url{https://lasp.colorado.edu/eve/data/} \\
SDO AIA Level 1 image data product: \url{https://aia.lmsal.com/} \\
TIMED-SEE Level 3 spectra data product: \url{https://lasp.colorado.edu/see/data/}\\
GOES-R XRS Level 2 science-quality flare summary data product and EUVS bandpasses and responsivities: \url{https://www.ncei.noaa.gov/products/goes-r-extreme-ultraviolet-xray-irradiance} \\


\appendix
\section{Data Selection and Flare-Class Balancing}\label{sec:data_selection_preamble}

 The prediction target is defined as the spectral irradiance matrix $\mathbf{I} \in \mathbb{R}^{N \times \lambda}$, where $N$ is the number of time samples and $N_{\lambda} = 1012$ is the number of wavelength bins of each modeled spectrum.

\subsection{Data Selection and Flare-Class Balancing}\label{sec:data_selection}
We implemented a sample-filtering strategy that preserves spectral outliers associated with rare and transient solar events, such as large X-class flares, so that the models are not exclusively biased toward quiet-Sun conditions and smaller flares. A key consideration in this dataset construction is the operational duty cycle of EVE’s MEGS-B channel, which under nominal operations is exposed for approximately three hours per day of continuous observing (beginning in August 2010), supplemented by brief 5-minute exposures during the remaining hours, in order to mitigate long-term detector degradation. During solar flare campaigns, MEGS-B switches to a longer daily exposure window, allowing the instrument to follow the complete evolution of the flare. The MEGS-A detector, by contrast, operated continuously at a 1-minute cadence until a capacitor short on May 26, 2014, after which no spectra below $\sim$33 nm were recorded. During their respective exposure windows, both instruments typically record spectra with an internal sampling cadence between 10 and 60 seconds, depending on the operation mode. For this work, we restricted our sample to times when both MEGS-A and MEGS-B observations were simultaneously available and resampled all spectra to a uniform 1-minute cadence. While MEGS-A alone yields over two million spectra from 2010 to 2014, the requirement of joint MEGS-A and MEGS-B coverage reduces the usable dataset to several hundred thousand spectra.

Although this constraint limits the number of samples, it has practical advantages. The reduced cadence of MEGS-B acts as a temporal filter, selecting a subset of the solar record that is more evenly distributed across different activity levels. The inclusion of flare stare mode observations from MEGS-B ensures better coverage of active periods and enhances the diversity of spectral conditions represented in the training set. As a result, the models trained under this selection strategy are expected to generalize more reliably across a wide range of solar activity.

To ensure that our training and test datasets adequately represent the full range of solar activity, spanning from quiescent sun to major flares, we leveraged the GOES X-ray flare data to construct a flare-class-balanced subset of the EVE observations. The flare list was obtained from the Level 2 GOES XRS flare summary product \texttt{sci\_xrsf-l2-flsum}, which provides a single consolidated record per flare, including its class and reported start, peak, and end times, spanning the period from 2010 to early 2020. For each flare in this list, we grouped events by their GOES flare class (X, M, or C) and determined approximate duration of each flare by identifying the time difference between its start and end status entries in the XRS flare summary dataset. This duration was used with a 10 minute temporal padding to the start and end time to define a temporal window around the central flare time, ensuring that the extracted samples include both the impulsive and decay phases of each event. For each event, the flare time was matched to the nearest timestamp in the EVE irradiance time series. We then retained only those intervals for which simultaneous MEGS-A and MEGS-B observations were available. This resulted in a set of 26 X-class, 384 M-class, and 4,425 C-class flares. The extracted spectra totaled approximately 700 (X-class), 8,500 (M-class), and 82,000 (C-class) samples.

To complement this flaring subset, we also constructed a non-flaring sample set by identifying all time indices not associated with any flare-related window. This non-flaring set consisted of approximately 208,000 samples, drawn from periods when the Sun was either quiet or exhibiting non-flaring activity. Because MEGS-B nominally collects only two spectra per day during non-flaring periods, these non-flaring intervals naturally span a wide range of quiet sun conditions, from deep solar minimum to non-flaring active regions. 

By enforcing this class-stratified sampling strategy, we mitigated the inherent bias toward quiescent conditions present in the raw dataset. Given the natural distribution of flare occurrences over the solar cycle, 33\% of the dataset consists of flare-associated samples (including X, M, and C classes) and 66\% consists of non-flaring intervals. This sampling ratio yielded the best overall training performance for both models, reflecting an optimal balance between flare class representation and sufficient training sample size.

We also evaluated a configuration with an equal number of samples drawn from each flare class ($\sim$700 samples from X, M, C) and $\sim$700 samples from non-flaring set. Although this fully balanced setup provided symmetry across classes, the limited number of available M and C class flares, including the quiet sun events, led to underfitting and significantly degraded model performance, even for X-class flares themselves. In this configuration, the hybrid model performed only marginally better than the proxy-only model, with improvements of less than $1.2\times$ in the relevant error metrics. Our exploratory analyses indicated that training on a broader set of C- and M-class flares, including non-flaring periods, contributes meaningfully to learning the spectral patterns associated with more extreme X-class events, likely due to shared underlying emission characteristics and continuum behavior.

\subsection{Dimensionality Reduction for Train-Test Splitting}\label{sec:data_train_test}

EUV spectra from EVE span more than a thousand wavelength bins, and the solar conditions that produce these spectra range from quiet-Sun emission to highly structured, rapidly evolving flare states. When building a train–test split for models that operate on the full spectrum, it is important to ensure that the training set includes examples from all of the major solar regimes. To distinguish common spectra from those that occur only a few times in the entire mission record, we used a compact representation of each spectrum, obtained by projecting it onto the leading spectral modes (the dominant patterns of variation across wavelength) using principal component analysis. This provides a method to identify spectra that deviate strongly from the bulk of the mission data. The spectra flagged in this step correspond to physically rare and anomalous situations: anomalous flare intervals, times when multiple active regions are simultaneously visible, or brief periods of intense high-temperature emission. These events are too few and too diverse to be divided between training and testing in a representative way; they do not form a coherent group from which one could hold out a subset without depriving the model of the information needed to learn their spectral structure. For this reason, all such rare but physically important spectra were included in the training set so that the model would encounter the full range of solar behavior, while the test set reflects typical daily variability without being dominated by isolated extreme events.

After generating the flare-class balanced dataset, we applied Principal Component Analysis (PCA) to the spectral irradiance matrix, $\mathbf{I}$, to reduce its dimensionality for train-test data splitting. The spectral irradiance at each time step was projected onto the top three principal components, resulting in a transformed matrix $\mathbf{I}_\mathrm{PCA} \in \mathbb{R}^{N \times 3}$. We then computed the Mahalanobis distance of each sample from the PCA mean vector $\boldsymbol{\mu}$, using the covariance matrix $\boldsymbol{\Sigma}$ of the PCA-reduced data:
\begin{equation}
D_i = \sqrt{ (\mathbf{I}_i - \boldsymbol{\mu})^\top \boldsymbol{\Sigma}^{-1} (\mathbf{I}_i - \boldsymbol{\mu}) },
\end{equation}
where $\mathbf{I}_i$ is the $i$-th row of $\mathbf{I}_\mathrm{PCA}$. A conservative threshold of $D_i > 3$ was used to identify statistical outliers. This yielded $~200{,}000$ inliers (non-outliers) and $8{,}000$ outliers out of $208{,}000$ total samples. The dimensional reconstruction error from PCA was negligible, with a mean squared error of $\sim 8.4 \times 10^{-12}$ between the reconstructed and original EVE spectra.

From the identified inliers, which span the full populated region of the PCA-reduced spectral space as quantified by the Mahalanobis distance, we randomly selected $80\%$ of the samples for the training set. This procedure ensures that spectra from all flare classes and activity levels are represented in both the training and test datasets. This training set was then augmented with all $8{,}000$ samples flagged as outliers, with the goal of ensuring that rare and anomalous spectral states, such as those arising during unusual flare phases or complex multi–active-region configurations, were represented in the training data. Here, the term ``outlier'' refers to spectra that are statistically rare in the reduced spectral space, not to entire flare classes; in particular, most X-class flare spectra fall within the inlier population and are therefore still subject to the same random train–test split. The final training set consisted of $168{,}000$ unique spectra, and the remaining $32{,}000$ samples formed the test set.

In addition to random sampling, we explicitly reserved several well-characterized solar events in their entirety in the test dataset. This was done to evaluate the models' ability to reproduce the full temporal evolution of both flaring and non-flaring conditions. Specifically, we manually selected and set aside the following time periods from the EVE dataset:
\begin{itemize}
  \setlength\itemsep{-0.25em}
    \item A quiet sun interval on 2010/05/16,
    \item An active region passage on 2012/05/16,
    \item An M-class flare event on 2011/02/14, and
    \item An X-class flare on 2012/03/07.
\end{itemize}
Each of these events was extracted with its full set of contiguous time steps, ensuring that their complete rise, peak, and decay phases, or in the case of the quiet sun and active region intervals, their temporal variability, were preserved. These event-based samples were explicitly excluded from the training set to prevent data leakage, thereby allowing for an unbiased assessment of the model’s capacity to generalize to unseen, temporally coherent events.

\section{Time-Resolved Model Performance for Spectral Lines with log T~$<$~6.3}\label{sec:cool_line_performance}
\begin{figure*}[h!]
    \centering
\includegraphics[width=0.90\textwidth]{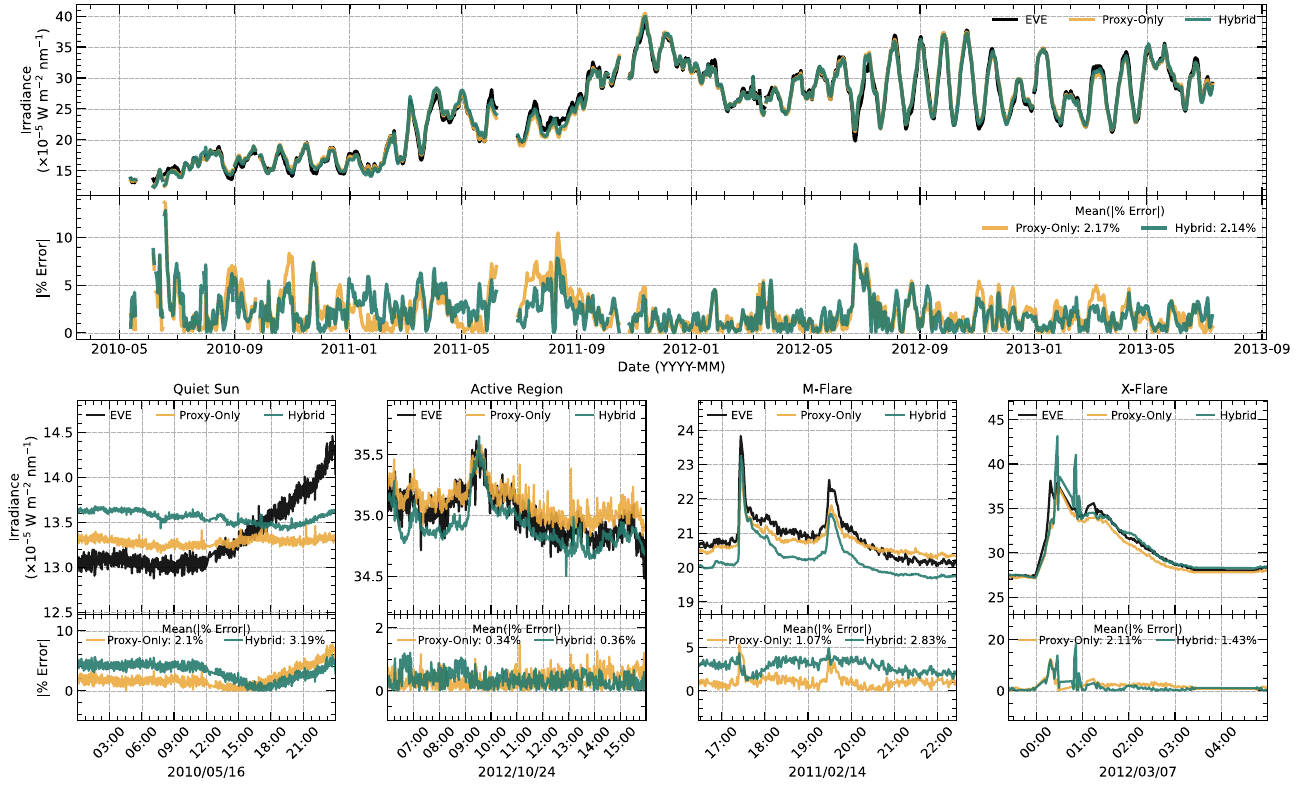}
    \vspace{-3mm}
    \caption{Same as Figure~\ref{fig:integrated_intensity_results} but for the Si~\textsc{xii}~(49.9~nm) spectral line (logT~$=$~6.29).} ~\label{fig:Si49.9_results}
\end{figure*}

\begin{figure*}[h!]
    \centering
\includegraphics[width=0.90\textwidth]{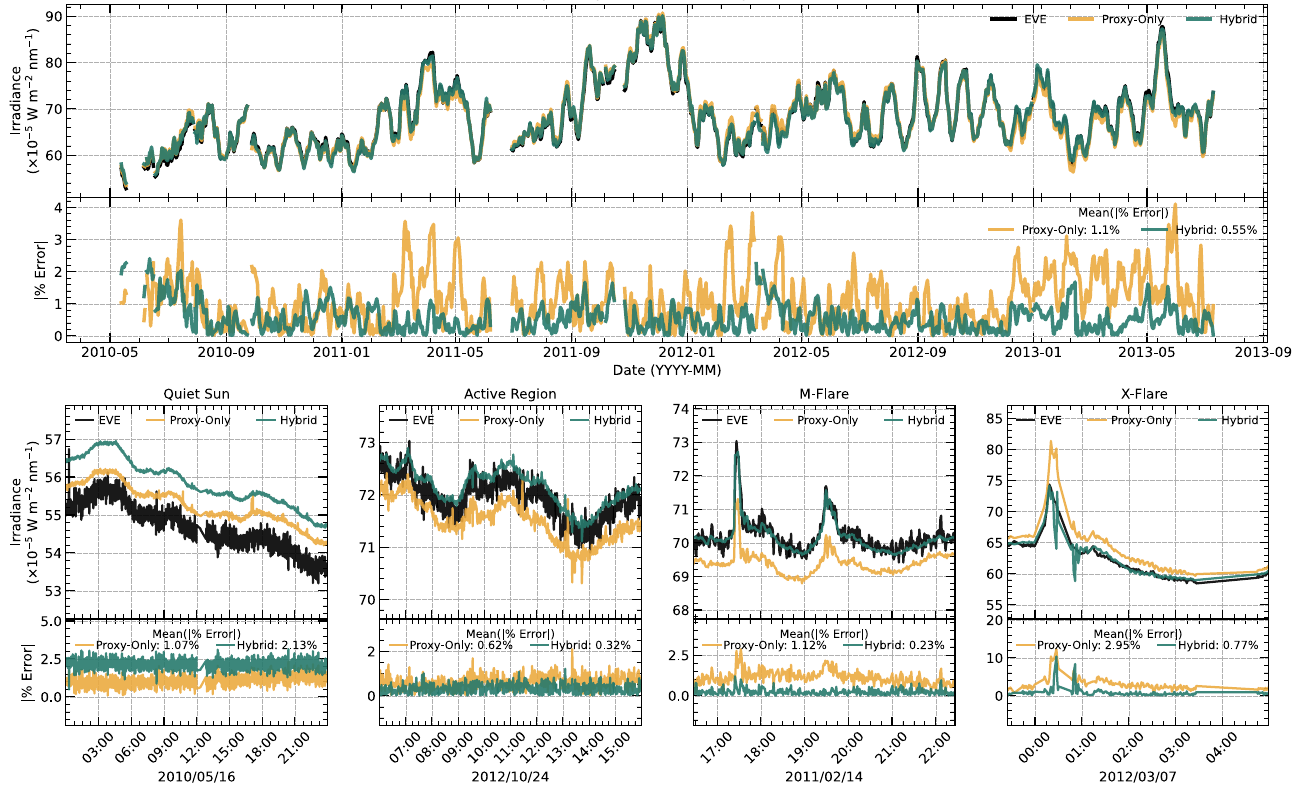}
    \vspace{-3mm}
    \caption{Same as Figure~\ref{fig:integrated_intensity_results} but for the Fe~\textsc{ix}~(17.1~nm) spectral line (logT~$=$~5.81).} ~\label{fig:Fe17.1_results}
\end{figure*}

\begin{figure*}[h!]
    \centering
\includegraphics[width=0.90\textwidth]{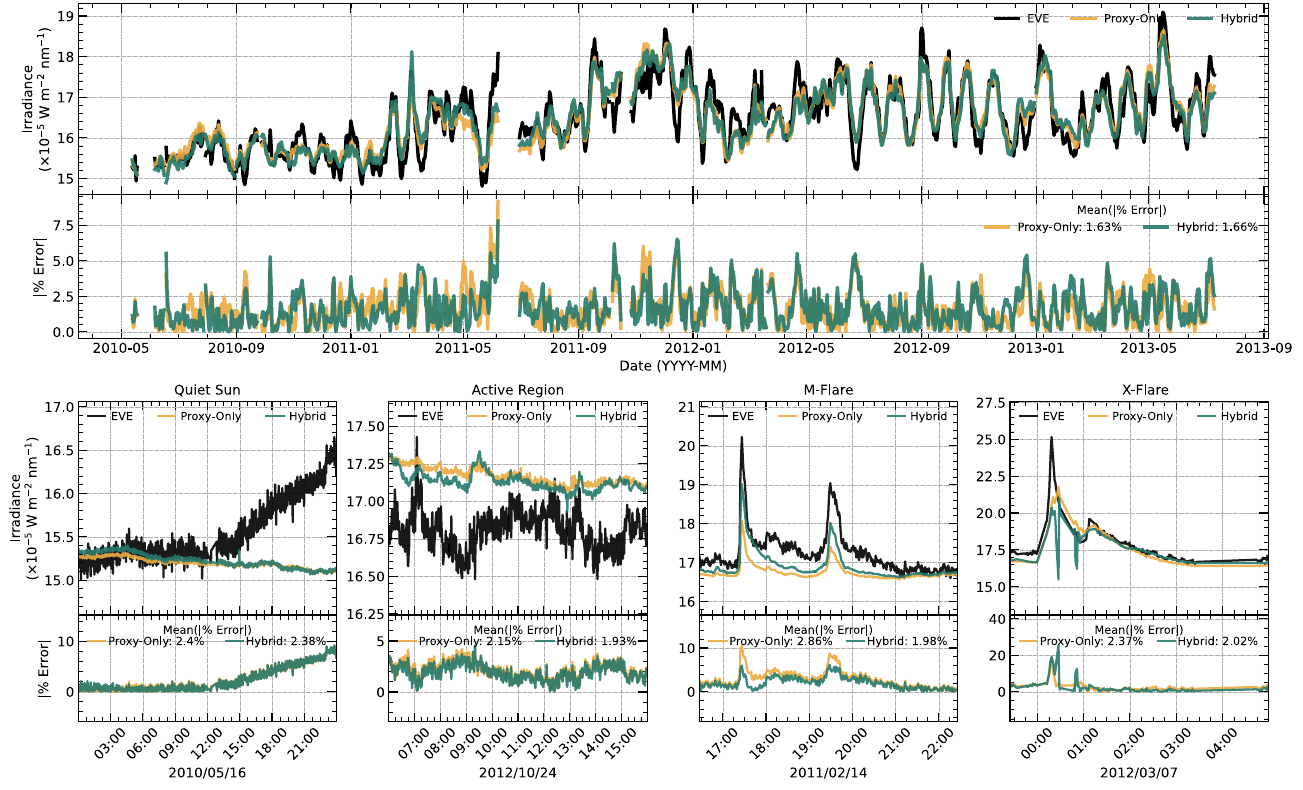}
    \vspace{-3mm}
    \caption{Same as Figure~\ref{fig:integrated_intensity_results} but for the Ne~\textsc{vii}~(46.5~nm) spectral line (logT~$=$~5.71).} ~\label{fig:Ne46.5_results}
\end{figure*}

\begin{figure*}[h!]
    \centering
\includegraphics[width=0.90\textwidth]{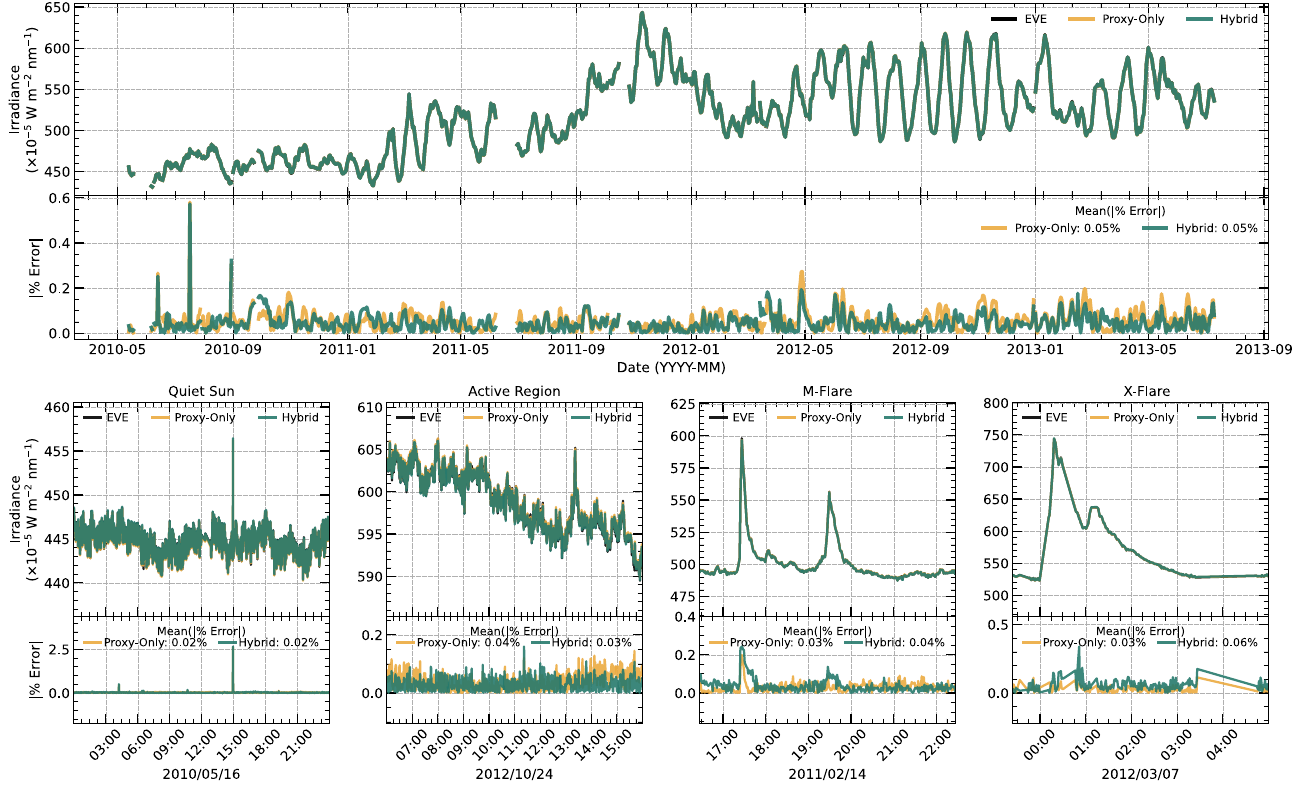}
    \vspace{-3mm}
    \caption{Same as Figure~\ref{fig:integrated_intensity_results} but for the He~\textsc{ii}~(30.4~nm) spectral line (logT~$=$~4.70).} ~\label{fig:He30.4_results}
\end{figure*}

The following figures show model performance in the same format as Figure~\ref{fig:integrated_intensity_results}, but for several spectral lines with log~T~$<$~6.3, sorted by formation temperature from high to low. Across this full set of additional spectral lines shown in this appendix, the hybrid model generally matches or outperforms the proxy-only model. In particular, the hybrid model shows improved performance for Si~\textsc{xii}~(49.9~nm) and Ne~\textsc{vii}~(46.5~nm) shown in Figure~\ref{fig:Si49.9_results} and Figure~\ref{fig:Ne46.5_results}, respectively, with similar results found for Fe~\textsc{xi}~(18.0~nm), Mg~\textsc{x}~(62.5~nm), Ne~\textsc{viii}~(77.0~nm), O~\textsc{iv}~(79.0~nm), O~\textsc{iii}~(52.6~nm and 83.5~nm), He~\textsc{ii}~(25.6~nm), and He~\textsc{i}~(58.4~nm). The hybrid model demonstrates superior performance for Fe~\textsc{ix}~(17.1~nm) shown in Figure~\ref{fig:Fe17.1_results} as well as for Fe~\textsc{xiv}~(21.1~nm) and Fe~\textsc{xiii}~(20.2~nm). For some lower-temperature lines, including  He~\textsc{ii}~(30.4~nm) in Figure~\ref{fig:He30.4_results}, as well as O~\textsc{v}~(63.0~nm) and He~\textsc{i}~(53.7~nm), the hybrid model performs comparably or marginally worse, consistent with the reduced sensitivity of DEM inversions under quiescent conditions. Overall, these results reinforce the conclusion that DEM-constrained spectral information provides substantial benefit over proxy-only models in regimes where rapidly evolving coronal plasma dominates the irradiance variability, while preserving comparable performance for lower-temperature continuum emission.


\clearpage




\bibliography{EUV_Modelling_Bib}{}
\bibliographystyle{aasjournal}



\end{document}